\documentclass[12pt]{article}
\usepackage[utf8]{inputenc}
\usepackage{graphicx}
\usepackage{amsmath}
\usepackage{xcolor}
\usepackage{rotating}
\usepackage{lscape}
\usepackage{longtable}
\usepackage{authblk}

\usepackage{lscape}
\usepackage{amssymb}
\usepackage{amsmath}
\graphicspath{{figs/}}

\usepackage[bottom=0.5in]{geometry}
\usepackage[square,comma,numbers]{natbib}

\newcommand{\aap}{Astronomy and Astrophysics}

\newcommand{\apjs}{The Astrophysical Journal Supplement Series}

\newcommand{\apjl}{The Astrophysical Journal Letters}

\begin{document}


\title{Estimating the confusion noise for the Millimetron space telescope.}


\author[1]{A.A. Ermash}
\author[1]{S.V. Pilipenko}
\author[1]{V.N. Lukash}
\affil{ {\it Lebedev Physical Institute, Russian Academy of Sciences, Leninskii pr. 53, Moscow, 119991 Russia}\\}

\date{\today}

\maketitle



\sloppypar
\vspace{2mm}
\noindent
   {Sensitivity of future far infrared 10m class space telescopes will be limited by a confusion noise created by distant galaxies.} 
   {Our primary goal is to create a model that will allow us to estimate the confusion noise parameters of the Millimetron mission.}
   {We construct a model of the Cosmic Infrared Background (CIB) aimed at exploring the methods of prediction and reduction of the confusion noise.
    The model is based on the publicly available eGALICS simulation.
    For each simulated galaxy we construct a spectral energy distribution with the help of the GRASIL code.
    To put our model in the context of the current CIB investigations, we compare the outputs of the model with the available
    observational data and with three other models.  
    One is a well known "backwards evolution" model of Bethermin et al. 2011 and two others are based on a simple mass-luminosity (M-L) relation applied to simulated dark matter halo catalogs.      
   }
   {We conclude that our model reproduces the observational data reasonably well.
    All four models show significant differences in the predictions of the distribution of sources on the flux-redshift plane, especially at high redshifts.
    We give estimations of the confusion noise based on number counts (source density criterion, probability of deflection criterion etc.) and based on the 
    analysis of the simulated maps.
    We show that resolution effects influence the number counts curves and noise estimations.
    }
   {}
 

{$^{*}$ A. A. Ermash: aermash@asc.rssi.ru}

\clearpage
\section*{\bf Introduction}
\renewcommand{\baselinestretch}{1.0}

 Our Universe is filled with background radiation in the full range of electromagnetic spectrum.
 Due to huge progress in the Far InfraRed (FIR) astronomy in the past few decades, the Cosmic Infrared Background (CIB) is getting more and more attention. 
 This background is created by the radiation of submillimeter galaxies which have maximum of their spectral energy distribution (SED) at about $100 \mathrm{\mu m}$ in the galaxy's rest frame due to large amounts of dust produced by the active star formation.
 The density of submillimeter galaxies on the sky is so high, that for FIR telescopes with apertures less than few tens of meters a fraction of these galaxies will be unresolved, i.e. there is a problem of confusion which affects sensitivity of FIR telescopes \citep{2004ApJS..154...93D}.
 
 The sensitivity of future space FIR telescopes such as Millimetron \citep{2012SPIE.8442E..4CS,2014PhyU...57.1199K}, Callisto, and OST in the mode of wide band photometry will be limited by confusion. 
 Several approaches have been proposed to improve the sensitivity beyond the confusion limit.
 Optical data can be used to get positions of possible submillimeter sources and then to model their approximate SEDs and subtract their contribution from FIR observations \citep{2015ApJ...798...91S}, see also FASTPHOT algorithm~\citep{2013ascl.soft02008B}.
 Another opportunity is to use submillimeter spectral lines that can give redshifts and other properties of unresolved galaxies. This information can be used to decrease the confusion limit by an order of magnitude \citep{Raymond10}.
 A modern multi-band source extraction software can use information from the shortest wavelengths, where the resolution is the best, to better decompose crowded image at longer wavelengths.
 This is implemented in, e.g., GETSOURCES \citep{Menshchikov,2013A&A...560A..63M,2017A&A...607A..64M}.
 Papers by \cite{2016MNRAS.462.1989A} and \cite{2014ApJ...780...75D} should be mentioned in this context. 
 Authors create the sample of dusty starforming galaxies on high redshift by analysing difference maps created from Hershel observations.
 
 The usefulness of these methods can be estimated by simulation of observations with future instruments.
 Existing observations cannot be used for this purpose: the data from space observatories (Herschel, Spitzer, Akari)
 and from single-dish ground telescopes on larger wavelengths ($850\mathrm{\mu m}$, $1100\mathrm{\mu m}$)
 suffer from the relatively low angular resolution and sensitivity in comparison with future observatories.
 The observations of ALMA have high enough angular resolution and sensitivity, but they have a very limited area on the sky and will not cover the shortest wavelengths, $\lambda<300\mathrm{\mu m}$.
 Thus, we need an accurate model that will predict the distribution of sources with different spectra on the sky over redshifts and luminosities which can be probed by future space telescopes.
 Building such a model and testing it against existing observational data and other models is the goal of this paper.
 In next papers we plan to use this model to test various approaches in beating the confusion limit.
 
 There is already a large number of papers in which various models of the CIB are developed. They can be divided into two major groups:
  \begin{itemize}
   \item backwards evolution (sometimes also called phenomenological, see e.g.~\cite{2013MNRAS.428.2529H})
   \item semi-analytical models
  \end{itemize}
 In the first group of models the population of galaxies is described by a luminosity function (LF) or a series of luminosity functions of several populations of galaxies with different spectra.
 These LFs are evolving with redshift, and this evolution is parameterized by some mathematical law.
 The parameters of the evolution are found by fitting the model to all existing observational data, including source counts and measurements of the LFs.
 Since the LF data at low redshifts are the most complete, these data defines the shape of the LF which is then evolved backwards in time to fit other data, such as source counts.
 This type of models has advantage of very accurate reproduction of the observational data.
 Due to this advantage we use the backwards evolution model by \cite{2011A&A...529A...4B} as a reference for comparison with our own model.
 This model is discussed in Sect.~\ref{sec:data_and_approach} in more detail.
 On the other hand, backward evolution models may lack predictive power in the wavelength range not probed by observations.
 Such models also usually do not take into account the large scale structure of the Universe and the hierarchical clustering of matter.
 
 Models of this type were developed in many papers,
 e.g.~\cite{2011MNRAS.418..176R}, 
      \cite{2003ApJ...585..617D}, 
      \cite{2001ApJ...556..562C}. 

 They differ by the number of galaxy populations (as a rule, this number is from two to five) and by the way in which the evolution is specified.
  
 The second group is represented by semi-analytical models, that are based on the evolution of dark matter (DM) halos, which is described by the halo mass function or numerical simulations.
 Then halos with given mass $M$ are assigned luminosity using mass-to-light ratio, and quantities of interest such as source counts, LFs, can be computed by assuming some spectral energy distribution.
 Examples of such models can be found
 in~\cite{2010MNRAS.405....2L}, 
    \cite{2015MNRAS.446.1784C}, 
    \cite{2008MNRAS.391..420S}, 
    \cite{2010MNRAS.405..705F}, 
    \cite{2015A&A...575A..32C}, 
    \cite{2015A&A...575A..33C}. 

 The most accurate approach to model the galaxy evolution is a cosmological hydrodynamical simulation (see, e.g. \cite{2009Natur.457..451D}).
 But such simulations are limited in volume and it is hard to make estimations of the EBL based on them, though such attempts were made.
 See, e.g. \cite{2012MNRAS.427.2866S} where $850\mathrm{\mu m}$ and $1100\mathrm{\mu m}$ number counts are reproduced fairly well.
 
 One should note that classification of CIB models presented here is simplified, and there are models that incorporate features of different types of models.
 
 Earlier we have developed a simple semi-numerical model in which
 luminosities were assigned to the simulated halos
 according to a mass-luminosity relation~\cite{2017AstL...43..644P}.
 We have found the parameters of the mass-luminosity relation by fitting observed source number counts.
 The model reproduced source counts with high enough precision in the range 100--2000~microns.
 It also has reproduced the angular power spectrum of CIB observed by Herschel \citep{2013ApJ...779...32V}.
 However, our model used a single SED for all galaxies. 
 
 In this paper we construct a model of extragalactic background light (EBL) using the eGALICS simulation from \cite{2015A&A...575A..32C} and \cite{2015A&A...575A..33C}.
 The publicly available eGALICS data contains dark matter halo parameters as well as properties of stellar and gas components. 
 We create a SED library with the GRASIL and CHE\_EVO program codes \citep{1998ApJ...509..103S} and assign each model galaxy its individual SED.
 Then we create a model lightcone and analyse properties of model survey.
 The main advantages of such an approach are:
 1) The usage of N-body simulations guaranties the most precise large scale structure, hierarchical clustering and dark matter halo parameters.
 2) The absence of free parameters tuned to fit the observations.
 3) Taking into account the complex evolution of SEDs of galaxies.

 It is not the first time when the GRASIL code is used to create a SED library for an EBL model. See, e.g. the following papers:
 \cite{2010MNRAS.405....2L}, 
 \cite{2008MNRAS.391..420S}, 
 \cite{2010MNRAS.405..705F}. 

 Considering the resolution effects in a model of extragalactic background is important if number counts are considered.
 Because of rather wide beam of modern telescopes, a significant part of detected sources are blended groups.
 Up to date many papers dedicated to this problem are available, see, e.g. pioneering studies \cite{2011ApJ...743..159H,2012MNRAS.424..951H,2013MNRAS.434.2572H}, and also \cite{2013MNRAS.432L..85H}.
 This effect is most prominent on large submillimeter wavelengths: $850\mathrm{\mu m}$ and $1100\mathrm{\mu m}$.
 Due to significance of this effect \cite{2013MNRAS.434.2572H} propose a new term ``Submillimeter source'' instead of ``Submillimeter galaxy''.

\section{Millimetron space telescope}
 \label{sec:millimetron}
 The primary goal of this paper is to create a simple model that will allow us to gauge the CIB parameters for the planning Millimetron space telescope.
 The detailed characteristics of the telescope and scientific payload can be found in~\cite{2017ARep...61..310K,2014PhyU...57.1199K,2012SPIE.8442E..4CS} and on the official website of the project\footnote{millimetron.ru}.
 Below we describe the parameters that are vital for this work.

 The Millimetron space telescope will have a 10-m diameter primary mirror that will be actively cooled to the temperature $4.5$~K.
 The spacecraft will be launched  to the orbit near L2 point of the Earth-Sun system.  
 Photometric observations will be carried out with LACS (Long wave Array Camera Spectrometer) and SACS (Short wave Array Camera Spectrometer) instruments. 
 
 Short wave matrix spectrometer (SACS) will consist of two main parts~-- the matrix photometer operating in the whole frequency range,
 which is divided into several sub-bands by dichroic beam splitter, and a matrix spectrometer, the spectral resolution of which will be determined by the input optical filter.
 A similar approach was used in the PACS receiver (http://www.cosmos.esa.int/web/herschel/science-instruments), successfully operated as a part of the Herschel Space Observatory,
 which, undoubtedly, will be used in the development of short wave matrix spectrometer for the Millimetron observatory.

 LACS is similar to the SPIRE receiver (http://www.cosmos.esa.int/web/herschel/science-instruments) that successfully operated as a part of the Herschel Space Observatory.
 Simultaneously, this receiver will be optimized for precise measurement of Syunaev-Zeldovich effect.
 The spectrometer whole frequency range from 100 GHz to 1 THz will be divided into 4 sub-bands.

\section{EBL models}
 \label{sec:data_and_approach}
 In this paper we consider the following models.
 The first one is the model created by the authors of this paper, previous version of which was described in detail in~\cite{2017AstL...43..644P}.
 In this paper we refer to is as P2017.
 
 As a second model we considered the IRGAL project and the model presented by~\cite{2011A&A...529A...4B}.
 Here we refer to it as BM, the Bethermin backward evolution Model.
 
 Finally, we used the eGALICS model. 
 The simulation is described in copious details in~\cite{2015A&A...575A..32C} and \cite{2015A&A...575A..33C}.
 We prepare two models based on this simulation, E1 and E2, described below. 
 Main results of this paper are based on E2 model.
 
 These models of the EBL are based on different cosmological simulations and thus have different cosmological parameters.
 For the P2017 model the Cosmology was as following:
 $\Omega_{\Lambda0} = 0.693$,
 $\Omega_{m0}       = 0.307$,
 $H_0               = 67.8 \mathrm{km\,s^{-1} Mpc^{-1}}$~\citep{2016MNRAS.457.4340K}.
 In the BM model the cosmology was:
 $\Omega_{\Lambda0} = 0.734$,
 $\Omega_{m0}       = 0.266$,
 $H_0               = 71.0 \mathrm{km\,s^{-1} Mpc^{-1}}$~\citep{2011ApJS..192...16L}.
 And the eGALICS simulation was based on the WMAP 3-yr cosmology:
 $\Omega_{\Lambda0} = 0.76$,
 $\Omega_{m0}       = 0.24$,
 $H_0               = 73 \mathrm{km\,s^{-1}Mpc^{-1}}$.
 
 For all the calculations the Python 2.7 language was used.

 Let us briefly describe the approach used in creating the model that was previously published in~\cite{2017AstL...43..644P}.
 Using the COSMOSIM database we have extracted all available redshift cuts in the Small Multidark Planck numerical model~\citep{2016MNRAS.457.4340K}. 
 The size of the cube was $40\mathrm{Mpc\,h^{-1}}$ and the angle of the cone was set to $1^\circ\times 1^\circ$. 
 
 The orientation of the axis of the cone was set such that any part of the cube contributes to the cone only once.
 Minimum and maximum redshifts were set to $z_{min}=0.30$ and $z_{max}=6.19$.
 Minimum and maximum mass of a DM halo in the simulation was, respectively, $M_{min}=3\times10^{10}M_{\odot}$, $M_{max}=2.56\times10^{14}M_{\odot}$,
 the total amount of halos $N=1285307$.
 Lensing was accounted for with the assumption of a point lens model.
 
 We utilized the following frequently used equation as a $M$ -- $L$ ratio:
 \begin{equation}
  L(M,z)=L_0(1+z)^\eta\log(M)\exp\left(-\dfrac{(\log(M)-\log(M_0))^2}{2\sigma_L^2}\right)
  \label{eq:pl_m-l}
 \end{equation}
 The parameters were as following:
 $\log(M_0)=12.6$, $\sigma_L^2=0.15$, $\eta=3.16$ when $z<2$ and $\eta=0$ when $z>2$, $L_0=5\times10^9L_{\odot}$ for the IR luminosity in the wavelength range 8 -- 1000$\mathrm{\mu m}$. 
 Averaged spectra of galaxies were taken from ~\cite{2010A&A...514A..67M}.

 \subsection{eGALICS data}\label{sec:egalics_data}
  To create a model of extragalactic background based on the simulations of the baryonic matter 
  we made use of the data of the eGALICS project.
  The detailed description of this model can be found in the following two papers:~\cite{2015A&A...575A..32C} and \cite{2015A&A...575A..33C}.
  Creators of the simulation call it `Semianalytical'.
  As the first step they created the simulation of the dark matter with the following parameters.
  Cosmology~--  WMAP-3yr, where $\Omega_m=0.24$, $\Omega_\Lambda=0.76$, $f_b=0.16$, $h=0.73$.
  Volume of the simulation $(100h^{-1})^3\simeq150$Mpc$^3$, number of particles $1024^3$, each with mass $m_p=8.593\times10^7M_{\odot}$, minimal halo mass $M_h^{min}=1.707\times10^9 M_{\odot}$.
  Then authors of the model added baryonic matter taking into account the formation of discs, pseudobulges, supernova feedback, AGN, hot halo, cooling processes etc.
  One of the key new elements of the model is the cold non-starforming gas reservoir.
  
\begin{figure*}[ht!]
   \centering
   \includegraphics[width=0.48\columnwidth]{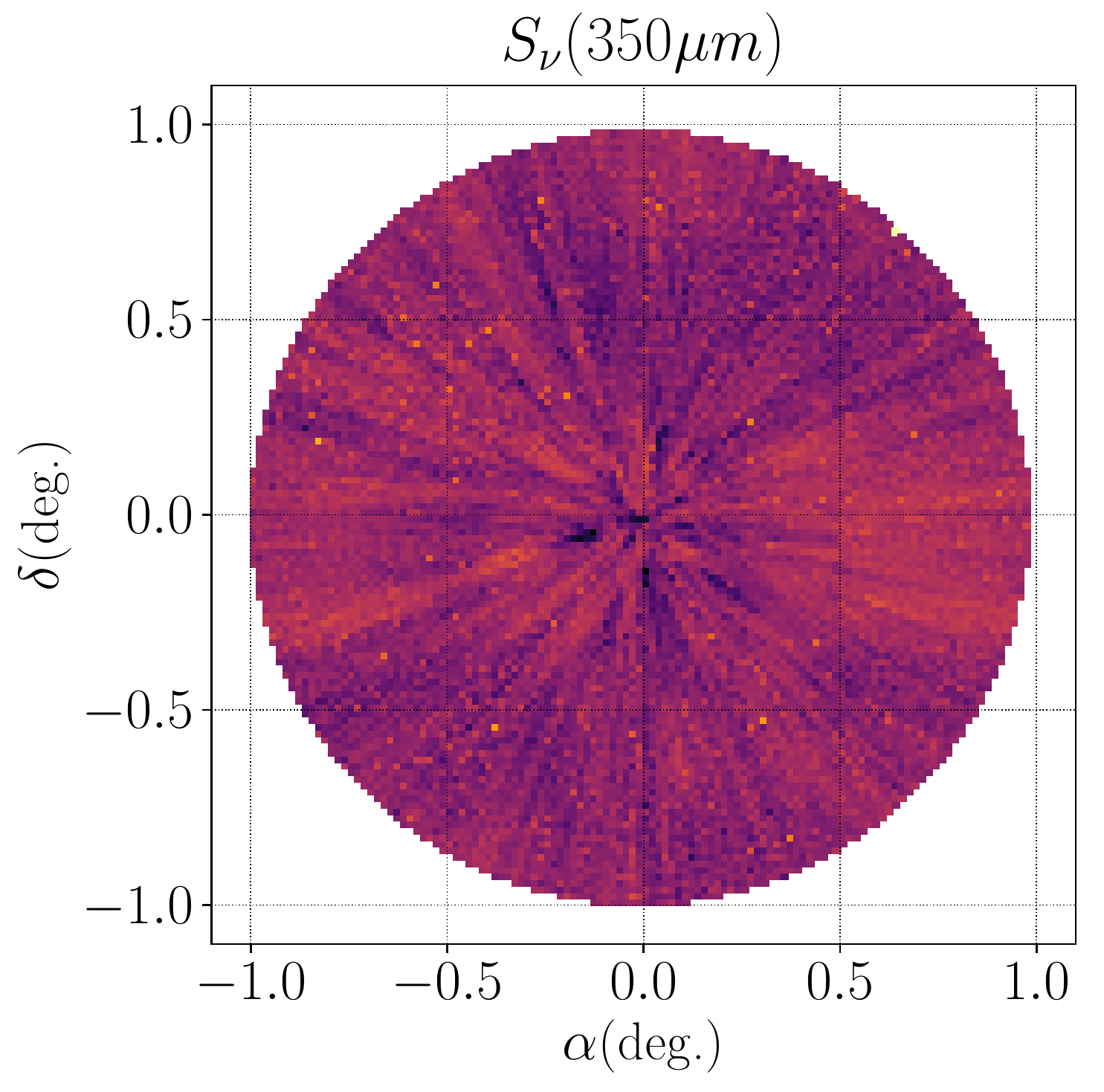}
   \includegraphics[width=0.48\columnwidth]{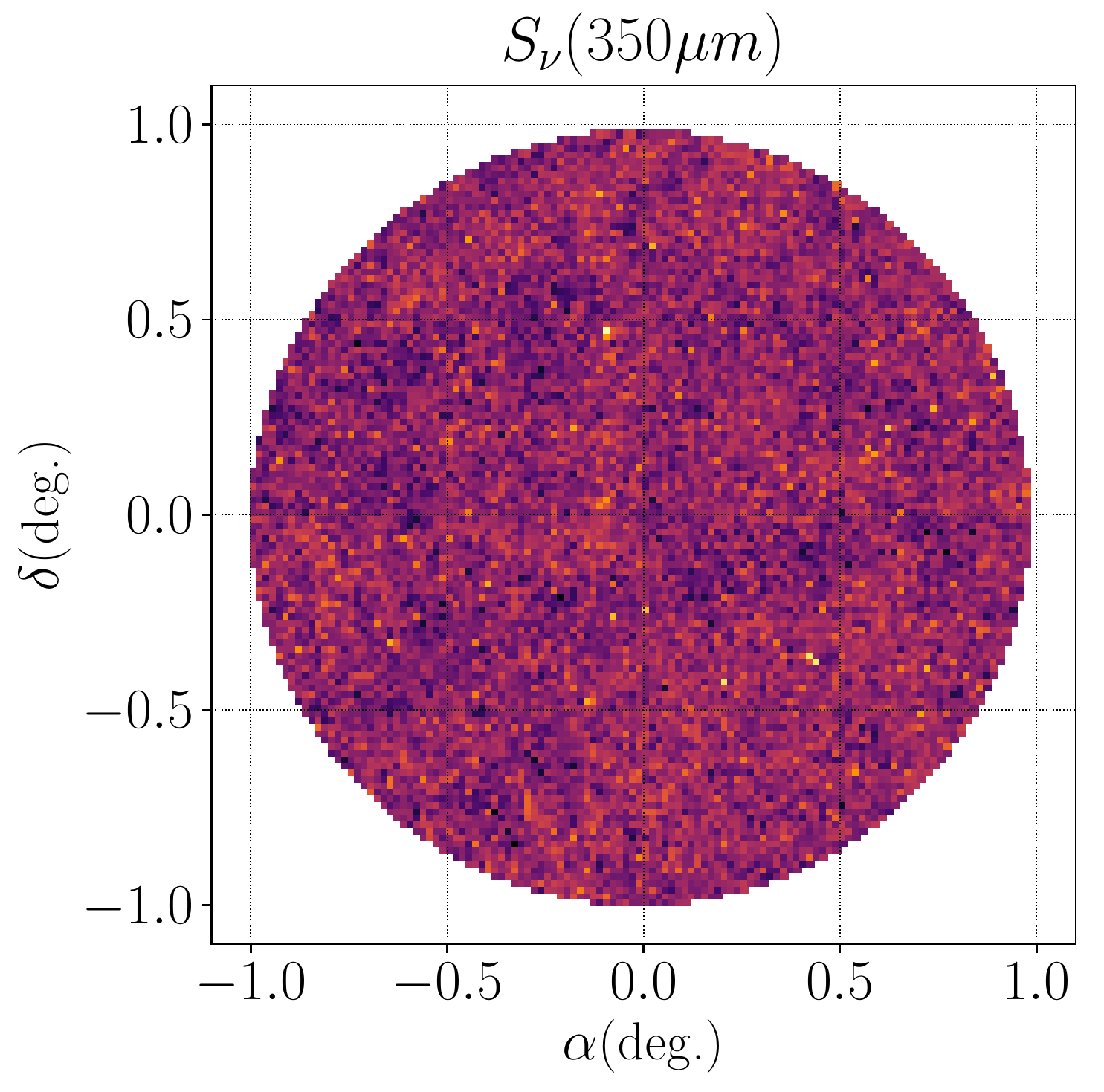}
   \caption{Model maps of the E2 model.   
   Wavelength 350$\mathrm{\mu m}$, pixel size is 1 angular minute.
   Left panel: maps without applying any transformation to the cubes. Repeated structures can be clearly seen.
   Right panel: transformations described in the text were applied to individual cubes. 
   The repeated structures can not be detected visually.}
   \label{fig:map_shift}
  \end{figure*}

  The first step one must take to create a background model from this simulation is to create a cone.
  If one creates a large cone from simulation with quite modest cube size a certain problem arises.
  The same part of the cube is included multiple times in the cone (see, e.g. \cite{2013MNRAS.429..556M}).
  Large scale structure evolves slowly, thus the repeating elements create the effect of perspective. 
  This effect is illustrated on the left panel of the Fig.~\ref{fig:map_shift}.
  Original solution to this problem was proposed in~\cite{2005MNRAS.360..159B}. 
  During the process of the creation of the cone each cube is affected by the following transformations independently on each axis:
  shift with random distance, rotation to $\pi/2$, $\pi$ or $-\pi/2$, and reflection along selected axis.  
  There are another approaches used in different models. E.g., \cite{2017A&A...607A..89B} rotated the box by $10^o$ along two coordinate axes.  
  The result of such transformation is shown in a Fig.~\ref{fig:map_shift}, right panel.
  As should be expected, the repeating structures are absent.
  The presence of such structures affects the outlook of the model map of the sky and the angular correlation function.
  Correlation function of the map without transformations shows excess clusterisation on low angular scales.

  We used the eGALICS data to create two models of EBL. The first is a simple one based on a $M$--$L$ relation, and the second took
  properties of baryonic matter into account. Comparison of these two models will show the importance of taking these properties into account. 
  
  At first let us consider a simple model of the EBL.
  We used information about dark matter halo mass, mass -- luminosity relation and a simple model spectrum.
  We refer to this model as E1.
  These ingredients are sufficient to calculate fluxes.
  The $M$--$L$ ratio was taken from~\cite{2014A&A...571A..30P}: 
  \begin{equation}
   \label{eq:f_p13_lm}
   L_{(1+z)\nu}(M,z)=L_0\Phi(z)\Sigma(M,z)\Theta[(1+z)\nu]
  \end{equation}
  where $\Phi(z)=(1+z)^\delta$ and  
  \begin{equation}
   \Sigma(M,z)=M\dfrac{1}{(2\pi\sigma^2_{L/M})^{1/2}}e^{-(\log_{10}(M)-\log_{10}(M_{eff}))^2/2\sigma^2_{L/M}}
  \end{equation}
  
  Parameters in this equation are, respectively:
  $\delta$ = 3.6,
  $\sigma^2_{L/M} = 0.5$,
  $\log_{10}M_{eff} = 12.6$ ($M_{eff}$ is in units of $M_{\odot}$),
  $L_0 = 0.0135 L_{\odot}$,
  $M_{min} = 1.0\times10^{10}M_{\odot}$. 
  First three parameters were adopted from~\cite{2014A&A...571A..30P}, the latter two from~\cite{2017MNRAS.467.4150W}.
  In case when $M<M_{min}$ luminosity $L=0$.
  If $z>2$ parameter $\delta=0$.
  
  We used the SED library of~\cite{2001ApJ...556..562C}. 
  The data is publicly available at the website:
  http://david.elbaz3.free.fr/astro\_codes/chary\_elbaz.html.
  The library consists of 105 spectra for the luminosity range $L=2.73\times10^8L_{\odot}$ -- $3.53\times10^{13}L_{\odot}$.
  As a simplistic first approximation such library allows to reproduce the variety of types of galaxies if we imply simple dependence of their parameters on mass.
  But it should be noted that usage of such a library is only justified if we aim at creating a simple model for illustrative purposes.
  The reason is that this library was created for local galaxies on low redshifts.
  The fact that SEDs of galaxies change with redshift was put under close consideration in numerous works. 
  See, for example, the paper dedicated to the GRASIL code:~\cite{1998ApJ...509..103S}.  
  
  Plots of differential number counts obtained by this approach are shown in
  Figs.~\ref{fig:diff_counts1} and \ref{fig:diff_counts2}.
  On these plots errors are Poissonian.
  Details see~\cite{1986ApJ...303..336G}.
   
  To obtain the results of interest from the eGALICS simulation we utilized the following approach.
  Hereafter we refer to this model as E2.
  Discs and bulges were treated independently due to the fact that in the simulation there is data available for them separately.
  We create a library of SEDs using the publicly available code GRASIL~\citep{1998ApJ...509..103S}.
  This code calculates spectral evolution of stellar systems taking dust effects into account.

  For our purposes two libraries of SEDs were created, for discs and bulges respectively. 
  The first one contained 16164 spectra, the second 7056 spectra.
  For each object in the eGALICS cone with disc and/or bulge the closest model in the parameter space was found.
  The parameters were:
  age of the galaxy $t_{gal}$, stellar mass $M_*$, mass of gas $M_{gas}$, star formation rate, and metallicity.
  For further processing we selected models that differ from the eGALICS object in stellar mass less than 0.1 dex.
  We also excluded models that are more than 5Gyr younger than the age of the Universe.
  The result of these operations is a model cone that contains three-dimensional coordinates of the objects and the identificator in
  of bulge and/or disc in the library.    
  
  To keep the calculation time reasonable, the step in each parameter should not be too small.
  To at least partly compensate the effects of the discrete step in each parameter we used the weighted sum of $N$ closest models in the parameter space.
  The $N$ value was set to $7$.
  
  During the next step of our work we had calculated the fluxes in the wavelengths of interest.
  This is necessary to calculate differential number counts, redshift distributions, estimate the confusion noise.
  
  It should be noted that for each bulge and disc we created 10 SEDs for the following inclination angles:
  $0^\circ$, $10^\circ$, ..., $90^\circ$.
  Each object was assigned a random inclination.
  
  Contribution of the Active Galactic Nuclei was also taken into account.
  The eGALICS data provide bolometric luminosities of active nuclei in galaxies.
  Bolometric correction to the IR was taken from~\cite{2004ASSL..308..187R}, where it has the following form $L_{IR}=0.19L_{bol}$.
  For simplicity's sake we used a single AGN SED template for AGN type 1 from~\cite{2017ApJ...841...76L}.  
  Their data reach only 2000$\mathrm{\mu m}$, but it is stated that on larger wavelengths the SED
  can be approximated by the black body SED with the temperature of 118~K.
  
  In our calculations we also accounted for lensing.
  The method implied was analogous to the one described above and implemented in P2017.
  Lensing affects submillimeter number counts due to their steepness, see e.g., ~\cite{2017A&A...607A..89B}. 
  In the following sections the results obtained with aforementioned model are described in detail.
  
   But there is one significant ingredient lost in such an approach because the properties of dust may change drastically with redshift.   
   \cite{2018ApJ...862...78C} and \cite{2018ApJ...862...77C}  consider two cases of dust evolution at $z=2$ and towards higher redshifts: dust rich or dust poor.
   Existing data about source counts cannot help to discriminate between these two scenarios.
   See, e.g. Fig.~5 in \cite{2018ApJ...862...78C}.
  In this paper this effect was not taken into account and was left for future work.
  
  Recently during the preparation of the current paper creators of the eGALICS model published two papers
  (\cite{2019arXiv190101906C} and \cite{2019arXiv190101747C})
  dedicated to the further development of their approach.
   In future we plan to use more advanced models of galaxy evolution to estimate the parameters of the extragalactic background.

\section{Results}
 \label{sec:results}

 \subsection{Number counts}
  \label{subsec:diff_counts}
  
  \begin{figure*}[ht!]
   \includegraphics[width=0.49\textwidth]{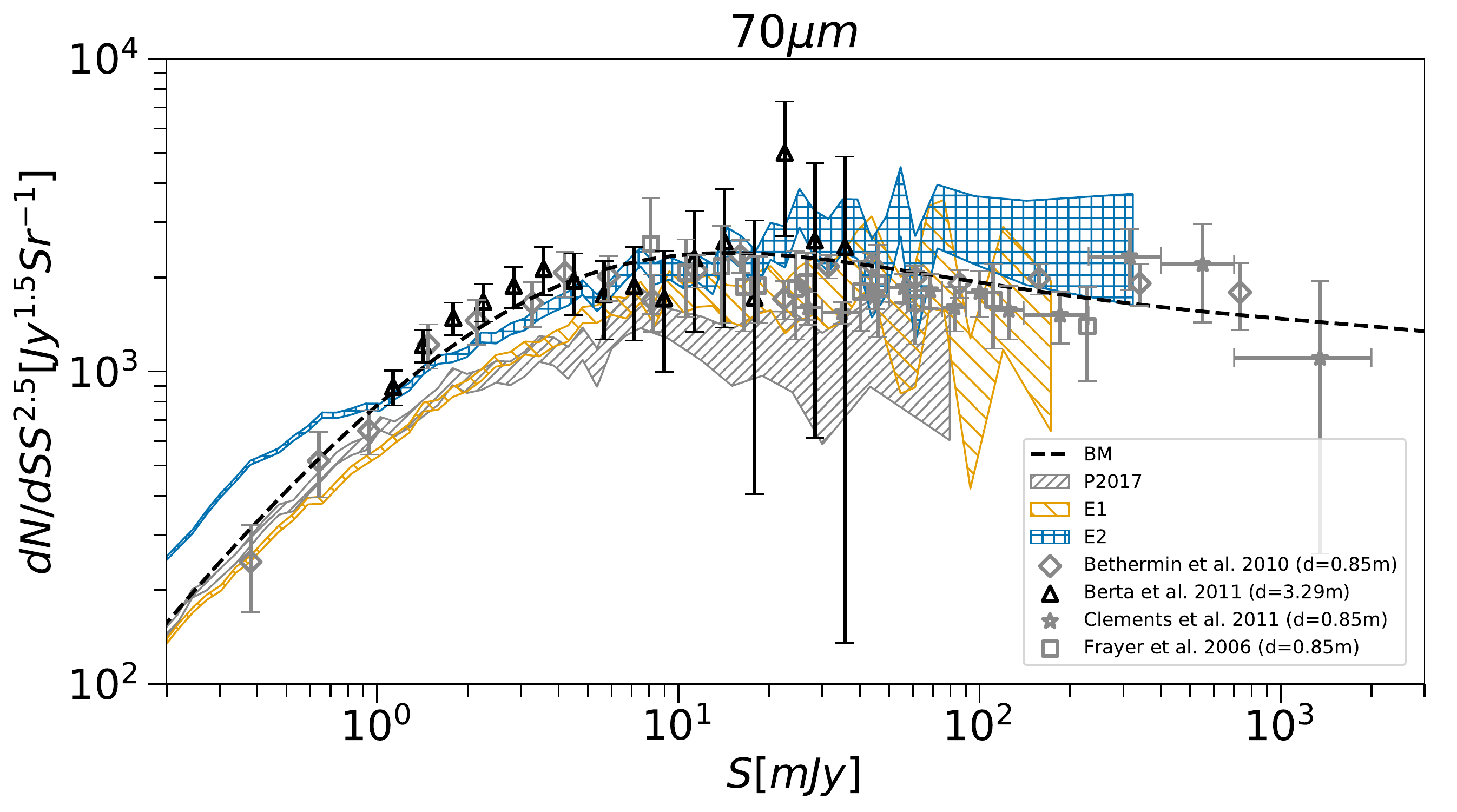}
   \includegraphics[width=0.49\textwidth]{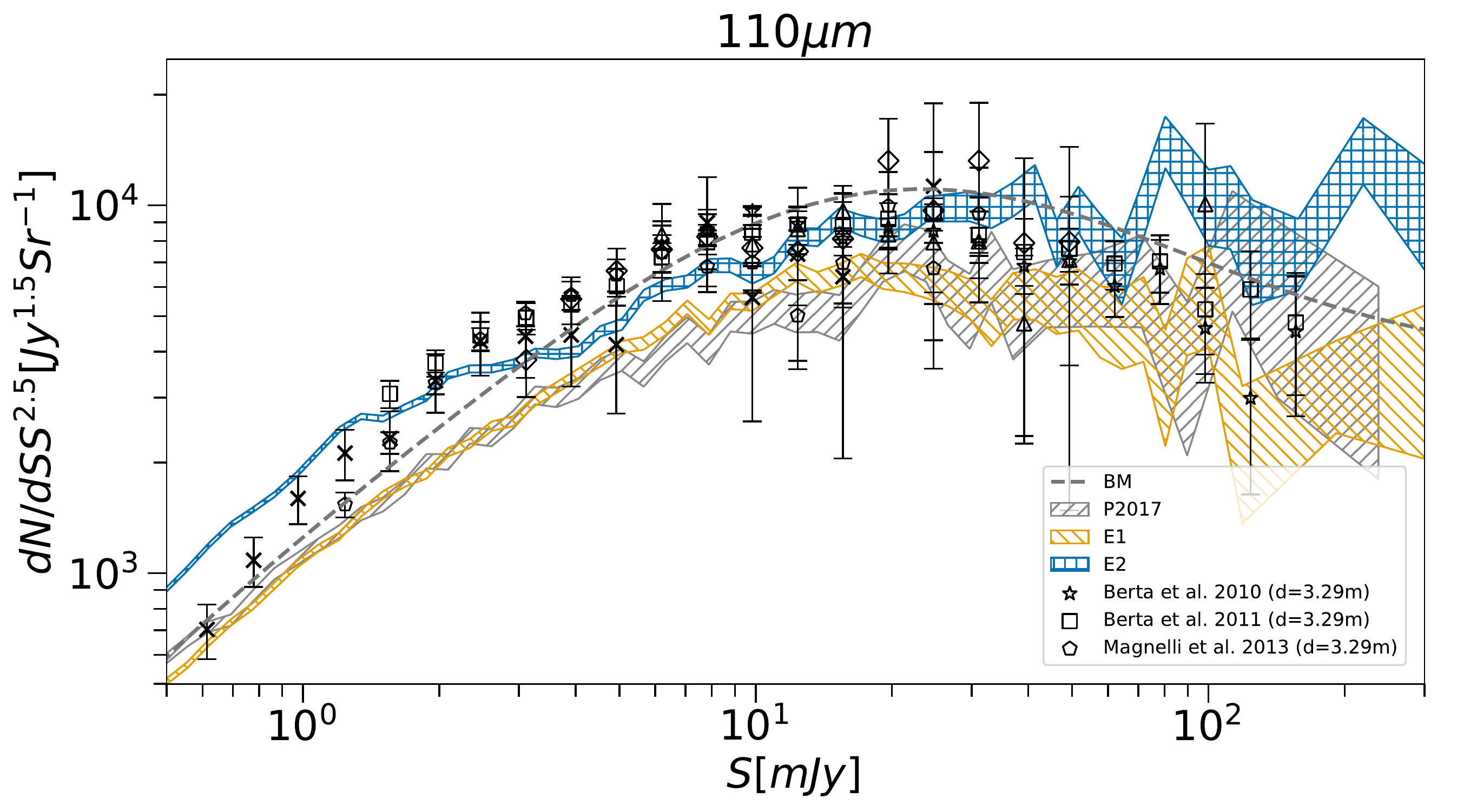}
   \includegraphics[width=0.49\textwidth]{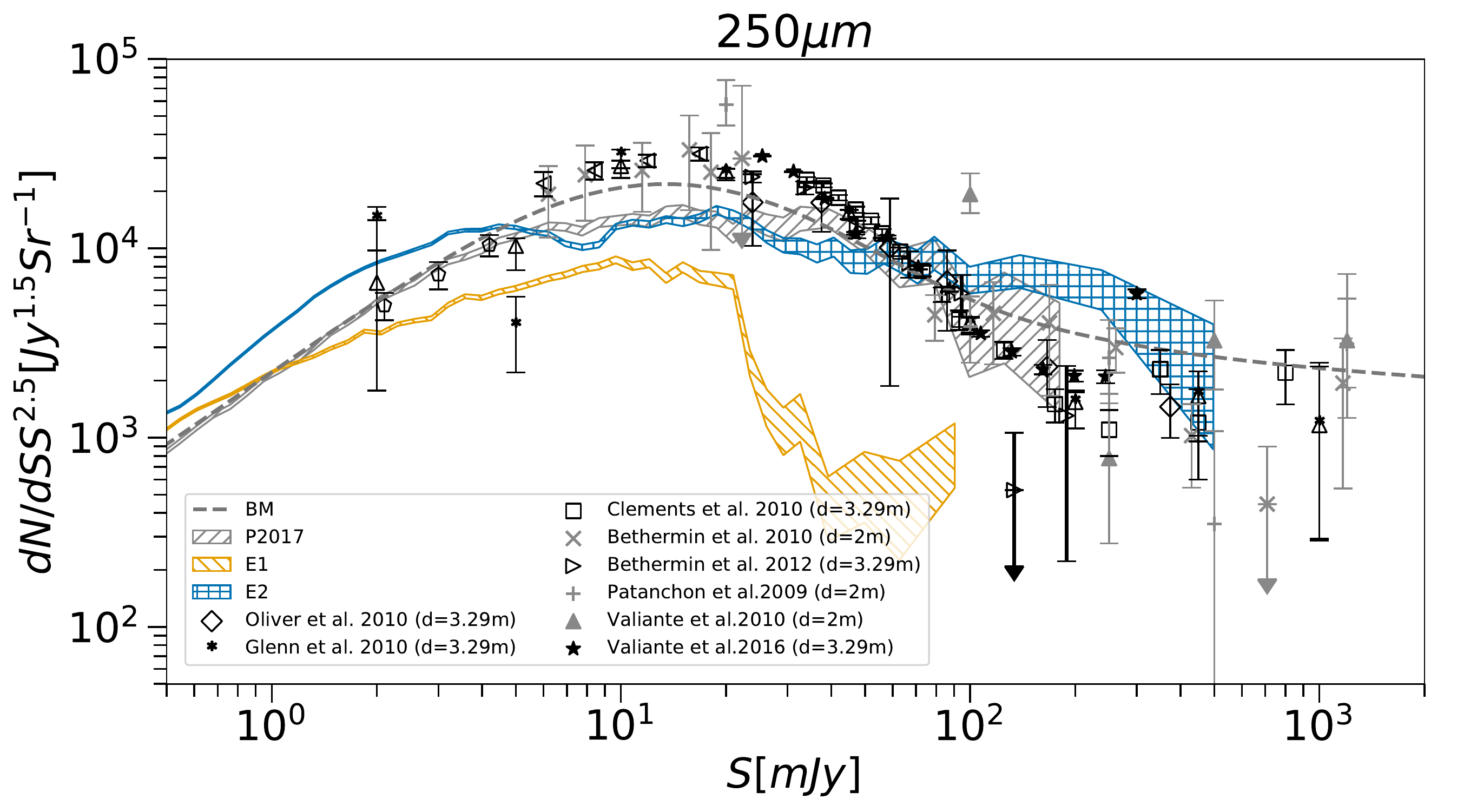}
   \includegraphics[width=0.49\textwidth]{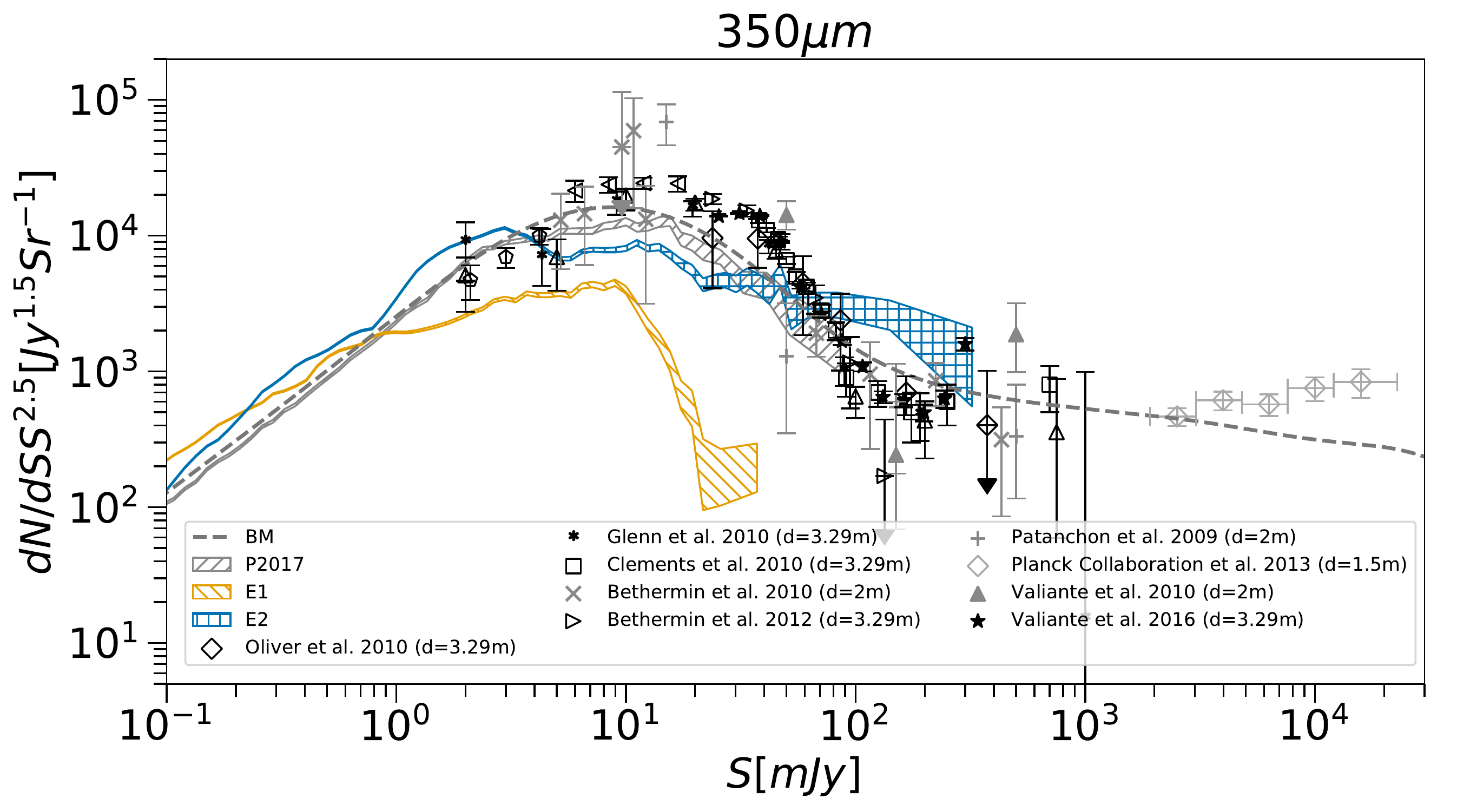}
   \caption{Differential source number counts.
    Common for all plots:
    1) Predictions of the BM model.
    2) The P2017 model.
    3) E1.
    4) E2.
    Various markers show data from the following papers.
    \textbf{Upper left ($70\mathrm{\mu m}$)}:
        \cite{2010A&A...512A..78B},                      
        \cite{2011MNRAS.411..373C},                      
        \cite{2006AJ....131..250F}~-- Spitzer $d=0.85$m. 
        \cite{2011A&A...532A..49B}~-- Hershel $d=3.29$m. 
    \textbf{Upper right ($110\mathrm{\mu m}$)}:
        \cite{2010A&A...518L..30B},                                          
        \cite{2011A&A...532A..49B},                                          
        \cite{2013A&A...553A.132M}~-- Hershel $d=3.29$m, $\lambda=100\mu$ m. 
    \textbf{Lower left ($250\mathrm{\mu m}$)}:           
        \cite{2010A&A...518L..21O}, 
        \cite{2010MNRAS.409..109G}, 
        \cite{2010A&A...518L...8C}, 
        \cite{2012A&A...542A..58B}, 
        \cite{2016MNRAS.462.3146V}~--       Herschel $d=3.29$m. 
        \cite{2010A&A...516A..43B}, 
        \cite{2009ApJ...707.1750P}, 
        \cite{2010ApJS..191..222V}~--       BLAST $d=2$m. 
    \textbf{Lower right ($350\mathrm{\mu m}$)}:
        Legend is the same as for 250$\mathrm{\mu m}$, with one exception: 
        \cite{2013A&A...550A.133P}~-- PLANCK $d=1.5\mathrm{m}$. 
    }
   \label{fig:diff_counts1}
  \end{figure*}
  
  \begin{figure*}[ht!]
   \includegraphics[width=0.49\textwidth]{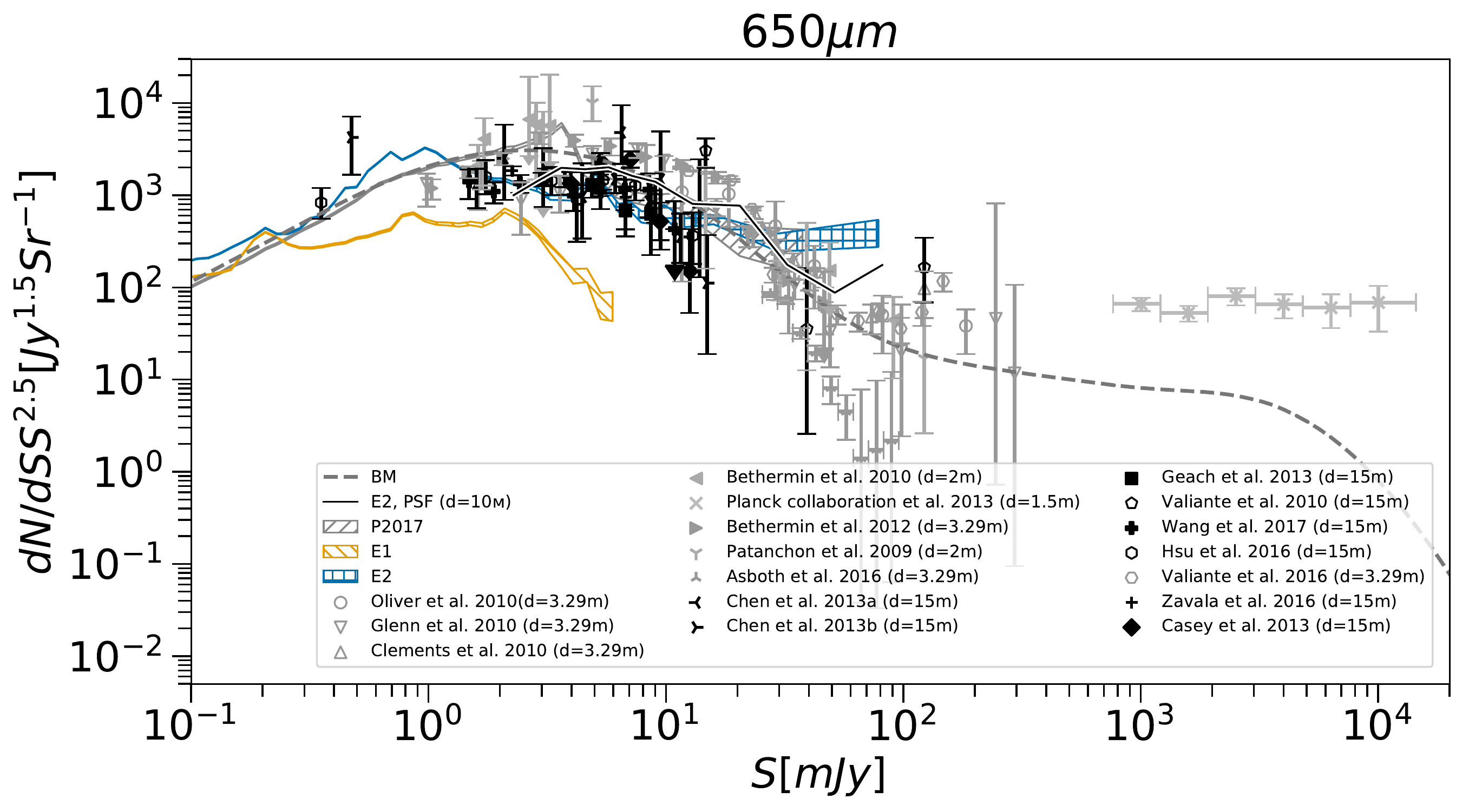}
   \includegraphics[width=0.49\textwidth]{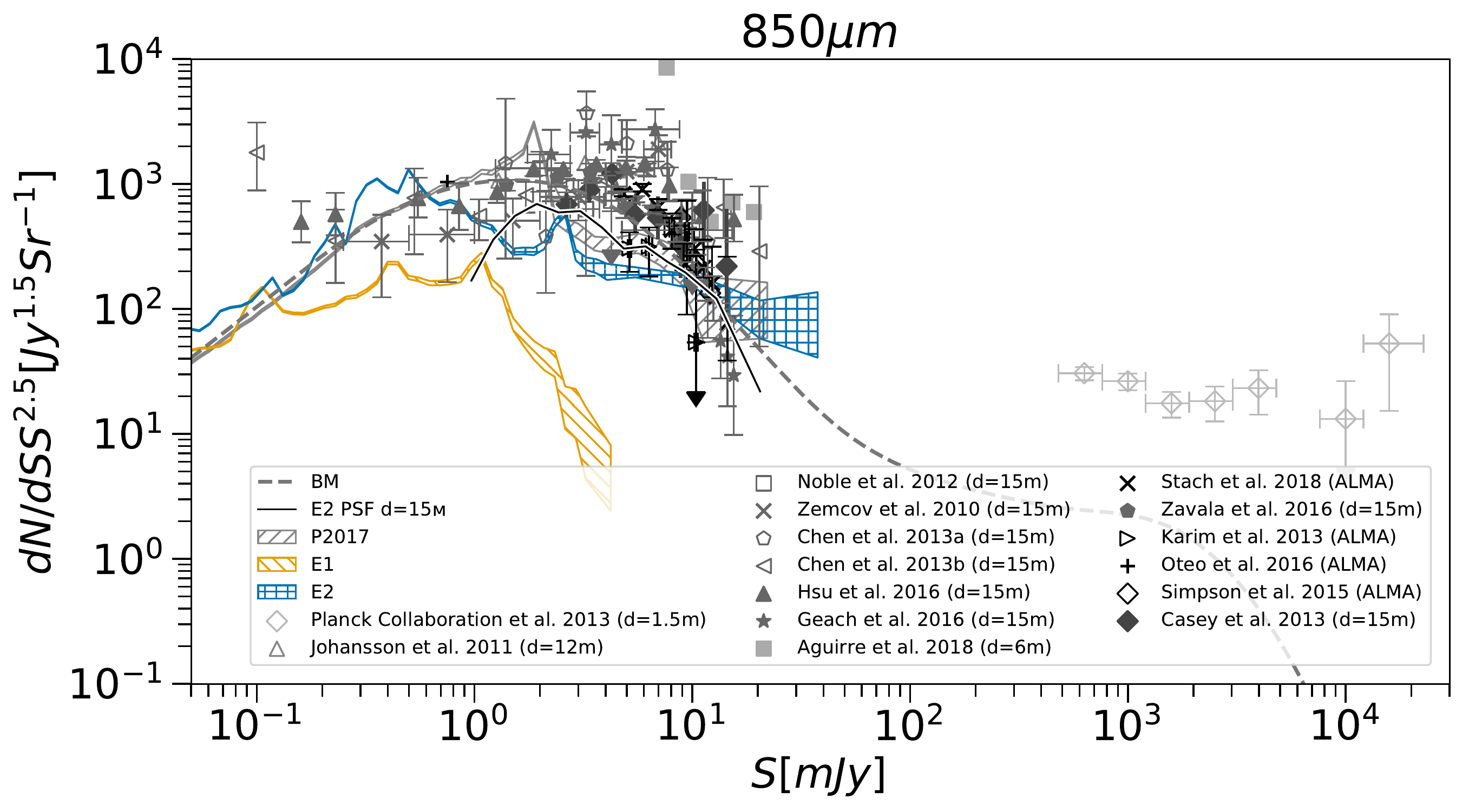}
   \includegraphics[width=0.49\textwidth]{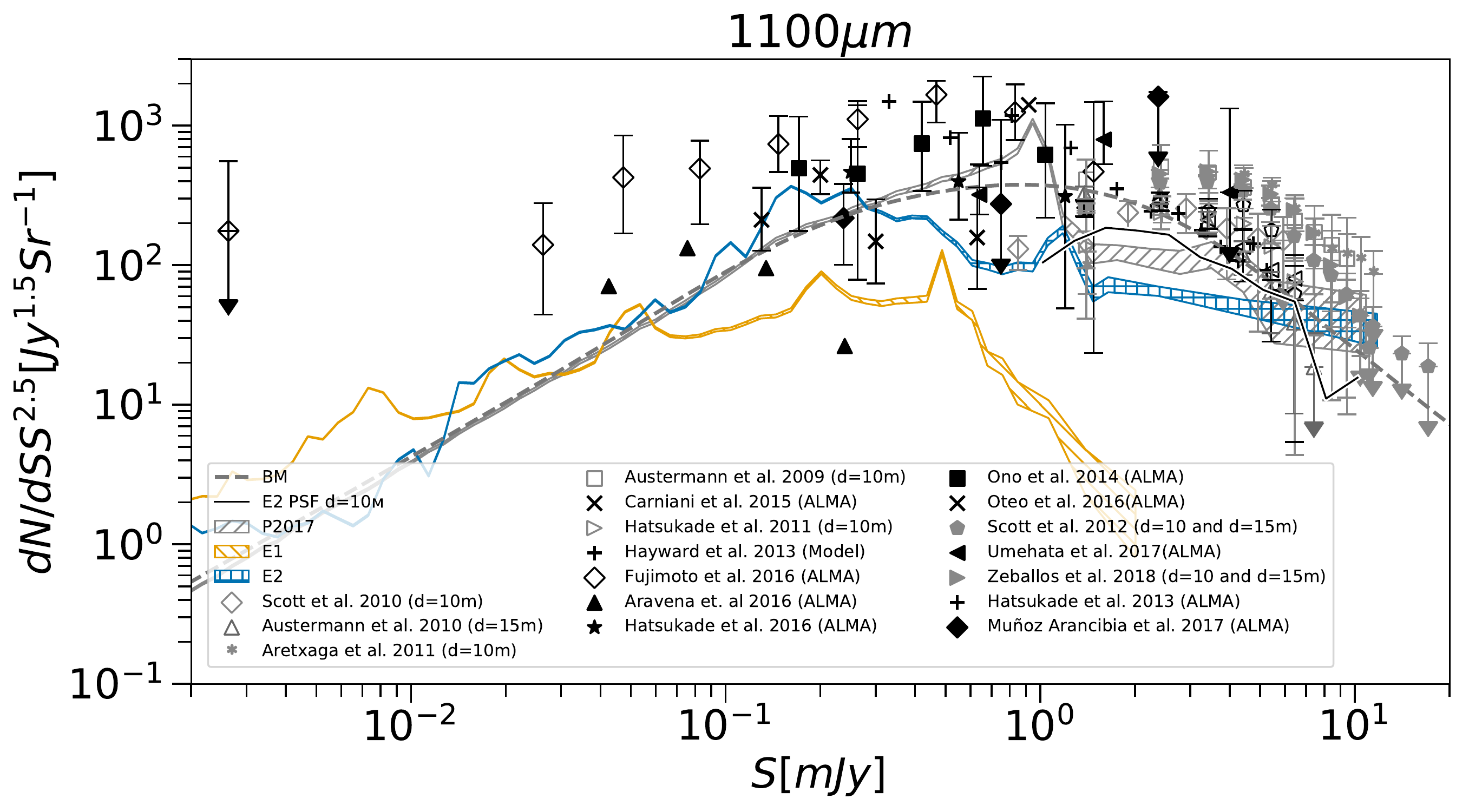}
   \includegraphics[width=0.49\textwidth]{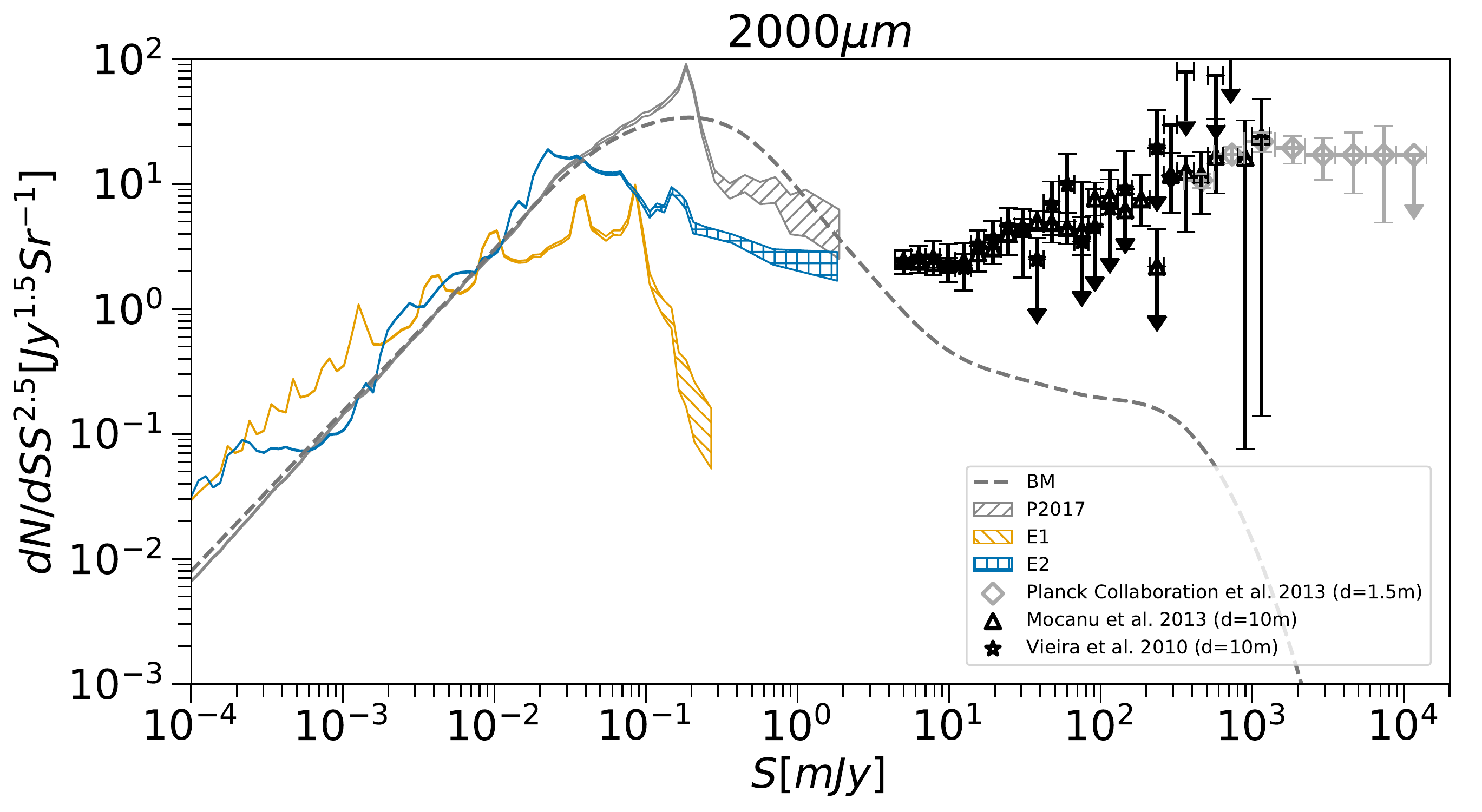}
   \caption{Differential source number counts.
    Common for all plots:
    1) Predictions of the BM model.
    2) The P2017 model.
    3) E1.
    4) E2.
    Various markers show data from the following papers.
    \textbf{Upper left ($650\mathrm{\mu m}$)}:
           \cite{2010A&A...518L..21O}, 
           \cite{2010MNRAS.409..109G}, 
           \cite{2010A&A...518L...8C}, 
           \cite{2012A&A...542A..58B}, 
           \cite{2016MNRAS.462.1989A}, 
           \cite{2016MNRAS.462.3146V}~-- Hershel $d=3.29$m, $\lambda=500\mu$m. 
           \cite{2013ApJ...762...81C}, 
           \cite{2013ApJ...776..131C}, 
           \cite{2013MNRAS.432...53G}, 
           \cite{2017ApJ...850...37W}, 
           \cite{2016ApJ...829...25H}, 
           \cite{2017MNRAS.464.3369Z}, 
           \cite{2013MNRAS.436.1919C}~--       SCUBA-2 JCMT $d=15$m, $\lambda=450\mu$ m. 
           \cite{2010ApJS..191..222V}, 
           \cite{2010A&A...516A..43B}, 
           \cite{2009ApJ...707.1750P}~--       BLAST $d=2$m, $\lambda=500\mu$m.    
           \cite{2013A&A...550A.133P}~--       Planck $d=1.5$m, $\lambda=550\mu$m. 
    \textbf{Upper right ($850\mathrm{\mu m}$)}:
           \cite{2013A&A...550A.133P}~-- PLANCK $d=1.5$m.                        
           \cite{2011A&A...527A.117J}~-- APEX LABOCA $d=12$m, $\lambda=870\mu$m. 
           \cite{2012MNRAS.419.1983N}, 
           \cite{2010ApJ...721..424Z}~-- SCUBA $d=15$m. 
           \cite{2013ApJ...762...81C}, 
           \cite{2013ApJ...776..131C}, 
           \cite{2016ApJ...829...25H}, 
           \cite{2017MNRAS.465.1789G}, 
           \cite{2017MNRAS.464.3369Z}, 
           \cite{2013MNRAS.436.1919C}~-- SCUBA 2 $d=15$m. 
           \cite{2013MNRAS.432....2K}, 
           \cite{2016ApJ...822...36O}, 
           \cite{2015ApJ...807..128S}, 
           \cite{2018ApJ...860..161S}~-- ALMA 870~$\mu$m. 
           \cite{2018ApJ...855...26A}~-- LABOCA/ACT $d=6$m, 870~$\mu$m. 
    \textbf{Lower left ($1100\mathrm{\mu m}$)}:
           \cite{2012MNRAS.423..575S}, 
           \cite{2010MNRAS.401..160A}, 
           \cite{2011MNRAS.415.3831A}, 
           \cite{2009MNRAS.393.1573A}, 
           \cite{2011MNRAS.411..102H}, 
           \cite{2018MNRAS.479.4577Z}~-- AzTEC $d=10$m, $d=15$m. 
           \cite{2013MNRAS.428.2529H}~-- model predictions. 
           \cite{2015A&A...584A..78C}, 
           \cite{2016PASJ...68...36H}, 
           \cite{2017arXiv171203983M}, 
           \cite{2017ApJ...835...98U}~-- ALMA. 
           \cite{2016ApJS..222....1F}, 
           \cite{2016ApJ...833...68A}, 
           \cite{2014ApJ...795....5O}, 
           \cite{2016ApJ...822...36O}~-- ALMA 1.2~mm. 
           \cite{2013ApJ...769L..27H}~-- ALMA 1.3~mm. 
           \cite{2011ApJ...737...83L}~-- IRAM 1.2~mm. 
    \textbf{Lower right ($2000\mathrm{\mu m}$)}:
           \cite{2013A&A...550A.133P}~-- PLANCK 143GHz ($2096\mathrm{\mu m}$). 
           \cite{2013ApJ...779...61M}, 
           \cite{2010ApJ...719..763V}~-- SPT. 
     On the 650, 850 and 1100$\mathrm{\mu m}$ panels number counts curves obtained by PSF photometry from simulated with the E2 model maps are shown.}
   \label{fig:diff_counts2}
  \end{figure*}
    
  The key prediction of any model of the extragalactic background is the differential number counts.
  The main product of the P2017 model as well as both E1 and E2 models are the cones containing coordinates, magnification coefficients due to the lensing
  and fluxes of   objects in the wavelengths of interest.
  Thus the calculation of number counts is trivial.
  
  The number counts were calculated for the following wavelengths:
  70, 110, 250, 350, 650, 1100, 2000$\mathrm{\mu m}$.
  Such a choice was made because this study focuses on the prediction of the background parameters for the Millimetron telescope.
  The number counts are shown in Fig.~\ref{fig:diff_counts1} and Fig.~\ref{fig:diff_counts2}.  
  It is important to compare model results with as much observational data as possible.
  
  If the observational wavelength differed from the wavelength of calculations, the observed number counts were converted the following way: $S_{\lambda_1}=k S_{\lambda_2}$.
  The $k$ coefficitient was obtained by linear regression of the fluxes of all the objects in the model cone.
  For 650$\mathrm{\mu m}$ there are no observational data, so all the points plotted show converted from 450, 500 and 550$\mathrm{\mu m}$ number counts.
  
  Many recent studies came to the same conclusion that AGN contribution is significant only at short wavelengths, see e.g. \cite{2018ApJ...862...77C}.
  In our case the influence of AGNs is only noticeable at 70 and 110$\mathrm{\mu m}$, at 250$\mathrm{\mu m}$ and longer wavelengths the contribution is negligible.
     
  At wavelengths larger than 1mm (1100$\mathrm{\mu m}$ and 2000$\mathrm{\mu m}$) AGN can contribute to number counts because of the synchrotron emission of radio-loud sources, but 
  because we consider a small sky area of one square degree the amount of such objects is negligible (\cite{2018ApJ...862...77C}, \cite{2011A&A...533A..57T}).  
  It should be kept in mind that errors in estimations of number counts on large fluxes are quite high because the area of simulated sky maps is limited.
   
  At high redshifts brightest 850 and 1100$\mathrm{\mu m}$ sources may represent a imhomogenous population that consists of mergers, unstable discs and other types of objects.
  In the simulation of barionic matter that our EBL model is based on mergings and their influence on star formation rate was included by it's creators.
  
  As could be expected, the phenomenological BM model approximates observational data on all considered wavelengths fairly well. 
  The P2017 predictions are within the error range of the observational data, but it predicts number counts to somewhat lower flux densities than the E2 model.
  The E1 model that was created for comparison shows significant deviation with the observations.
  The E2 model reproduces number counts reasonably well.
          
  An important fact is the discrepancy between observations and model predictions at $\lambda\geq650\mathrm{\mu m}$.   
  Such an effect was discussed, e.g. in \cite{2017A&A...607A..89B},
  \cite{2015MNRAS.446.1784C} (dedicated to the 850$\mathrm{\mu m}$ band)
  and, especially, \cite{2017MNRAS.469.3396C}.
  Originally this effect was discussed in \cite{2011ApJ...743..159H} and \cite{2012MNRAS.424..951H}.
  On long wavelengths there is a significant difference between the number counts obtained with single dish instruments and from model catalogs.
  The reason of it is that due to limited angular resolution merged sources are observed and such objects do not necessary belong to the same physical group (see, e.g., \cite{2015MNRAS.446.1784C} and also \cite{2013MNRAS.434.2572H}).
  This discrepancy is significant despite the high dispersion in observational estimates~\citep{2018ApJ...862...77C}.
  To illustrate this effect we created the model maps for all wavelengths considered.
  The resolution of the ``Millimetron'' was supposed to be limited by diffraction, pixel size in each wavelength was set to $FWHM/3$.
  For simplicity's sake we did not add any noise. It leads to the better source detection than in the real observations and, therefore, the distortion of number counts is created only
  by source confusion.
  Model maps were processed by the DAOPHOT algorithm that performs PSF-photometry.
  Corresponding number counts are plotted in Fig.~\ref{fig:diff_counts2}.
  
  Observational data in Figs.~\ref{fig:diff_counts1} and~\ref{fig:diff_counts2} are marked with different colors depending on the diameter of the main 
  mirror of the telescope.
  The general rule is that the darker the color, the higher the angular resolution.
  Such markers start to occupy different areas on the plot starting from 650$\mathrm{\mu m}$.
  It is important to note that number counts calculated from model catalogs better correspond to observational number counts obtained with high angular 
  resolution and vice versa~-- number counts from model maps better correspond to data from telescopes with lower aperture.   
   
  The shape of recovered sources number counts curve on lower fluxes can be explained by photometrical incompleteness.
  See, e.g., Fig.~2 from \cite{2016ApJ...832...78I}.   
  Correct predictions of submillimeter number counts is a challenging test for any model of extragalactic background, see, e.g. \cite{2016MNRAS.462.3854L}.      
  Future ``Millimetron'' telescope will help determine the true shape of the number counts curves on large wavelengths due to it's high angular resolution.

 \subsection{Redshift distribution}\label{subsec:redshift_distribution}
    
\begin{figure*}[ht!]
   \includegraphics[width=\textwidth]{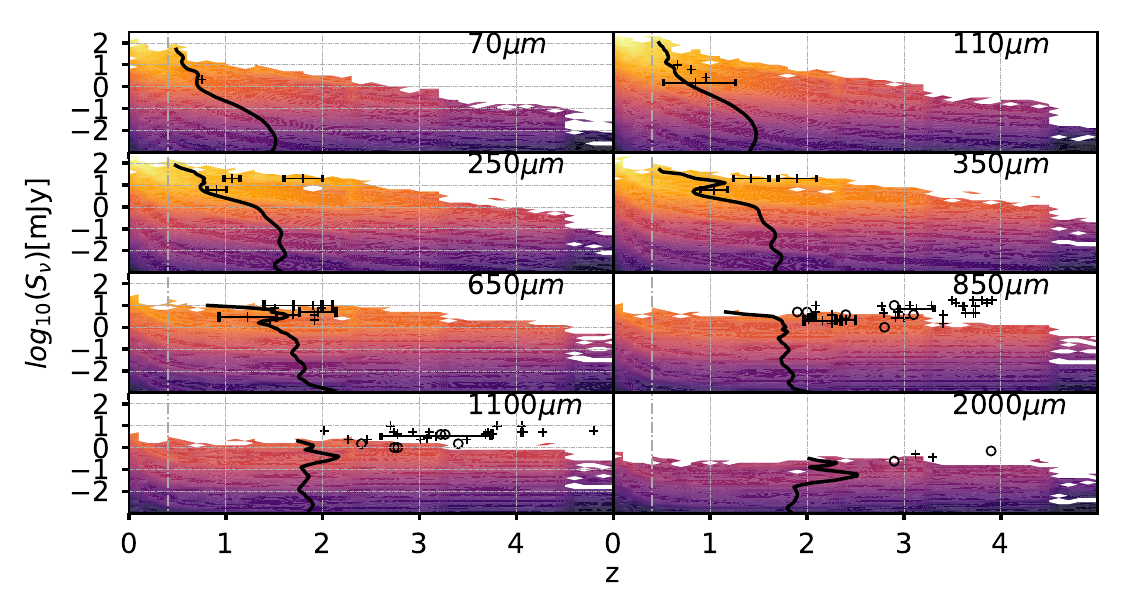}
   \caption{   
   Contribution to differential number counts ($dN/dS S^{2.5}$) from sources with different fluxes $S_\nu$ at different redshifts.
   The curve shows the dependence of average redshift of sources with flux greater than certain value $\langle z\rangle(S>S_{lim})$.
   Observational data shows the same. 
   Pluses show observational measurements, circles show model estimates.
     70$\mathrm{\mu m}$)                 
       \cite{2011A&A...532A..49B}~-- 
       Hershel.
     110$\mathrm{\mu m}$)      
       \cite{2013A&A...553A.132M}, 
       \cite{2011A&A...532A..49B}~-- Hershel. 
     250$\mathrm{\mu m}$)
       \cite{2012A&A...542A..58B}, 
       \cite{2012ApJ...753...23M}~-- Hershel. 
     350$\mathrm{\mu m}$)
       \cite{2012A&A...542A..58B}, 
       \cite{2012ApJ...753...23M}~-- Hershel. 
     650$\mathrm{\mu m}$)
       \cite{2012A&A...542A..58B}, 
       \cite{2012ApJ...753...23M}-- Hershel 500$\mathrm{\mu m}$. 
       \cite{2013MNRAS.436.1919C}, 
       \cite{2013MNRAS.436..430R}~-- (SCUBA-2 450$\mathrm{\mu m}$). 
     850$\mathrm{\mu m}$)
       \cite{2017A&A...607A..89B}, 
       \cite{2018ApJ...862...77C}, 
       \cite{2012MNRAS.427.2866S}, 
       \cite{2015MNRAS.446.1784C}~-- Model predictions.  
       \cite{2005ApJ...622..772C}~-- SCUBA.   
       \cite{2017MNRAS.471.2453S}, 
       \cite{2013MNRAS.436.1919C}, 
       \cite{2017MNRAS.469..492M}~-- SCUBA 2. 
       \cite{2014MNRAS.444..117K}~-- SMA, 890$\mathrm{\mu m}$. 
       \cite{2014MNRAS.444..117K}, 
       \cite{2017ApJ...840...78D}~-- Laboca 870$\mathrm{\mu m}$. 
     1100$\mathrm{\mu m}$)
       \cite{2018ApJ...862...77C}, 
       \cite{2013MNRAS.428.2529H}~-- Model predictions. 
       \cite{2009MNRAS.398.1793C}, 
       \cite{2014MNRAS.444..117K}, 
       \cite{2015A&A...577A..29M}~-- AzTec.  
       \cite{2014MNRAS.444..117K}~-- PdBI 1300$\mathrm{\mu m}$.  
       \cite{2017A&A...608A..15B}~-- ALMA 1250$\mathrm{\mu m}$.  
      2000$\mathrm{\mu m}$)
       \cite{2015A&A...576L...9B}, 
       \cite{2018ApJ...862...77C}~-- Model predictions. 
       \cite{2014ApJ...790...77S}~-- GISMO.             
   }
   \label{fig:2d_E2_twocol}
  \end{figure*}

  It is a well known fact that on different wavelengths the dependence of contribution to the source
  counts on redshift varies significantly. 
  Here we present 2d plots that show the dependence of contribution to number counts from objects with different
  fluxes on different redshifts for four models considered, see Fig.~\ref{fig:2d_E2_twocol}.
  
  The line on each plot shows the dependence of average redshift of sources with fluxes greater that certain value on flux $\langle z\rangle(S>S_{lim})$.
  To exclude the influence of small amount of bright local sources averaging was performed for objects with $z>0.4$.
  
  The general trend is the increase of contribution of high redshift sources with increase of wavelength.
  It can be explained by the negative K-correction (see, e.g., \cite{2014ApJ...780...75D}).
  For example, at 870$\mathrm{\mu m}$ the flux does not depend on redshift in the $z=1$--$10$ interval and is defined only by physical parameters of the
  galaxy~\citep{2013MNRAS.434.2572H,2013MNRAS.432L..85H}.

  At 1100$\mathrm{\mu m}$ and 2000$\mathrm{\mu m}$ bright sources have significantly higher redshifts that faint ones \citep{2018ApJ...862...78C}.
  So it can be said that the redshift distribution of a survey depends on the depth of the survey~\citep{2017A&A...608A..15B}. 
  All the above said leads to an important conclusion that to search for high redshift objects on $\lambda\geq850\mathrm{\mu m}$ we do not
  need deep surveys but shallow ones with large coverage area.
  
  The E2 model reproduces the redshift distribution up to 650$\mathrm{\mu m}$ fairly well.
  But on higher wavelengths the brightest objects have significantly higher redshift.
  Firstly, the high variance of observational estimates should be noted.
  So, at 850$\mathrm{\mu m}$ estimations for the brightest objects give $z=2$--$4$ and $z=2$--$5$ for 1100$\mathrm{\mu m}$.
  The result of our work will not be influenced by parameters of several brightest sources on the modeled sky area because our primary goal is
  the estimation of the confusion noise.
 
 \subsection{Extragalactic Background Light}
 \label{subsec:ebl_sed} 
 
  \begin{figure*}[ht!]
   \centering
   \includegraphics[width=0.49\columnwidth]{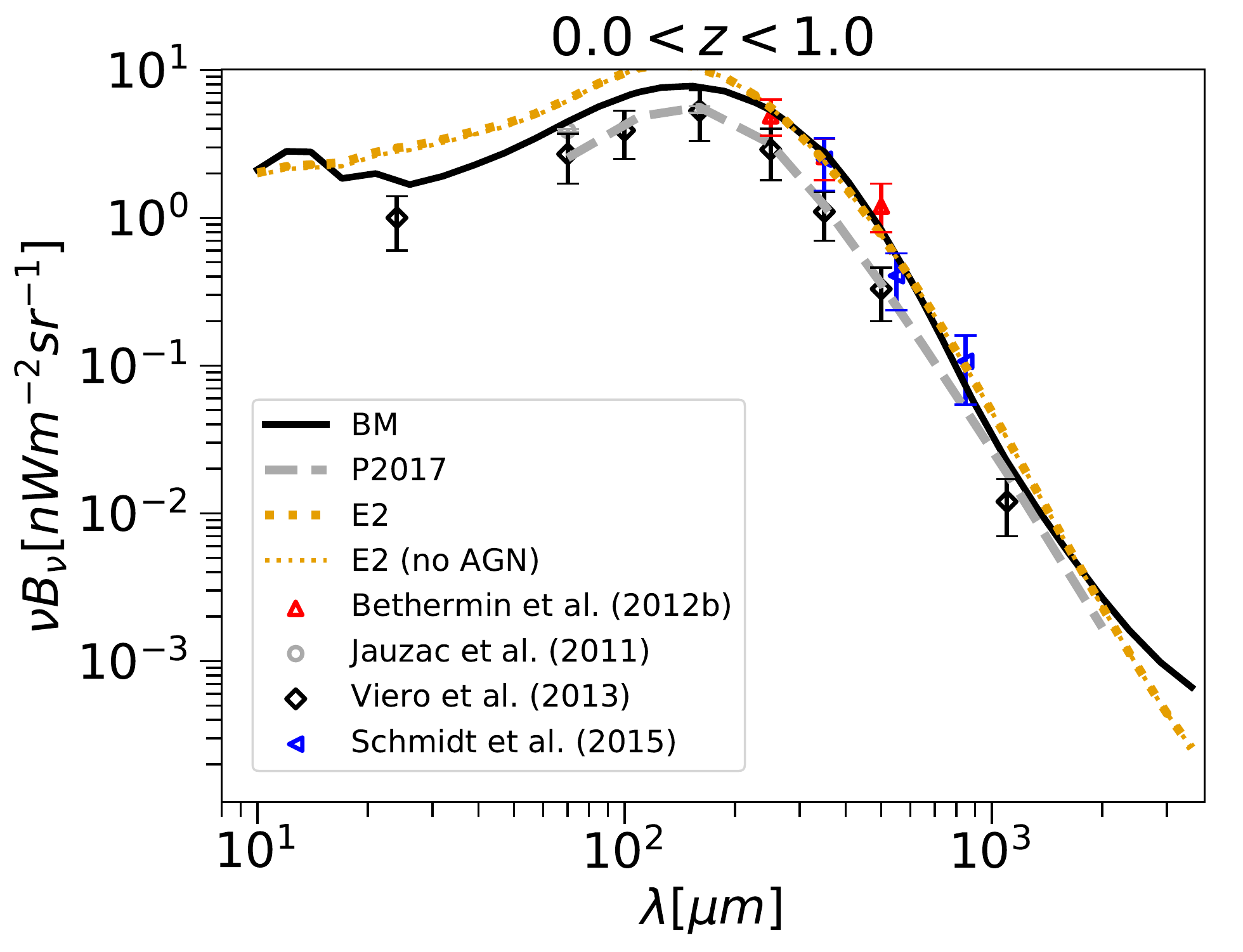}
   \includegraphics[width=0.49\columnwidth]{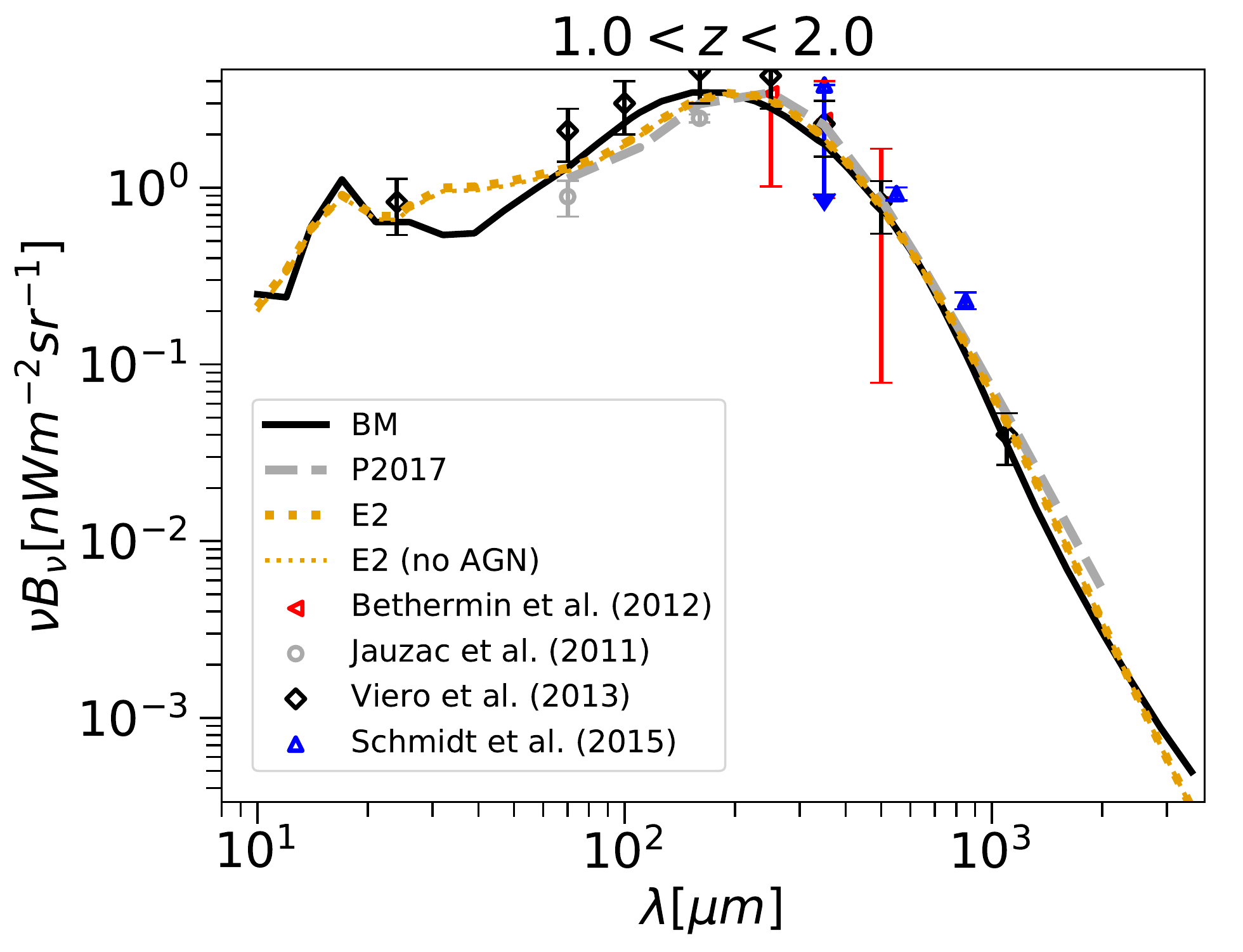}\\
   \includegraphics[width=0.49\columnwidth]{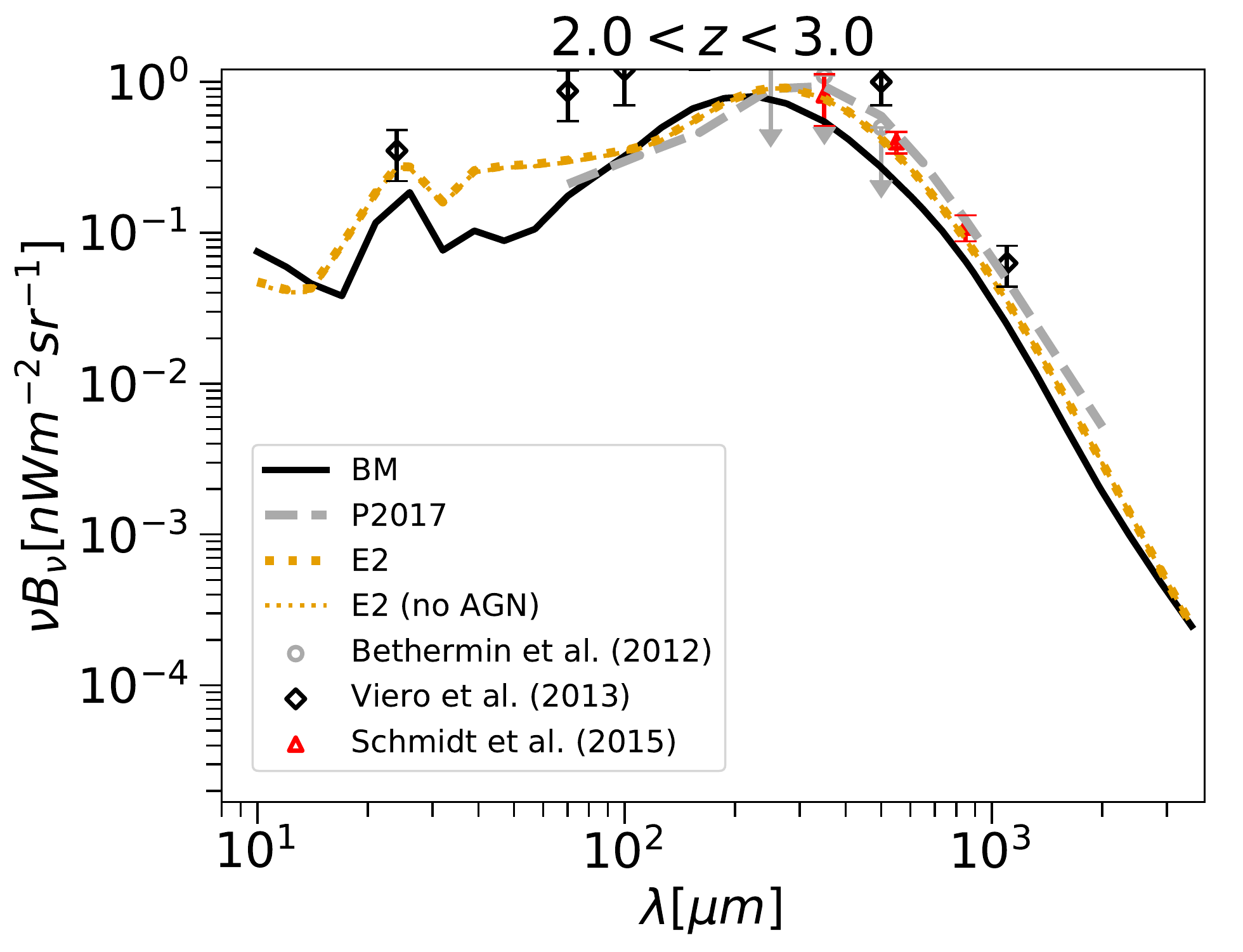}
   \includegraphics[width=0.49\columnwidth]{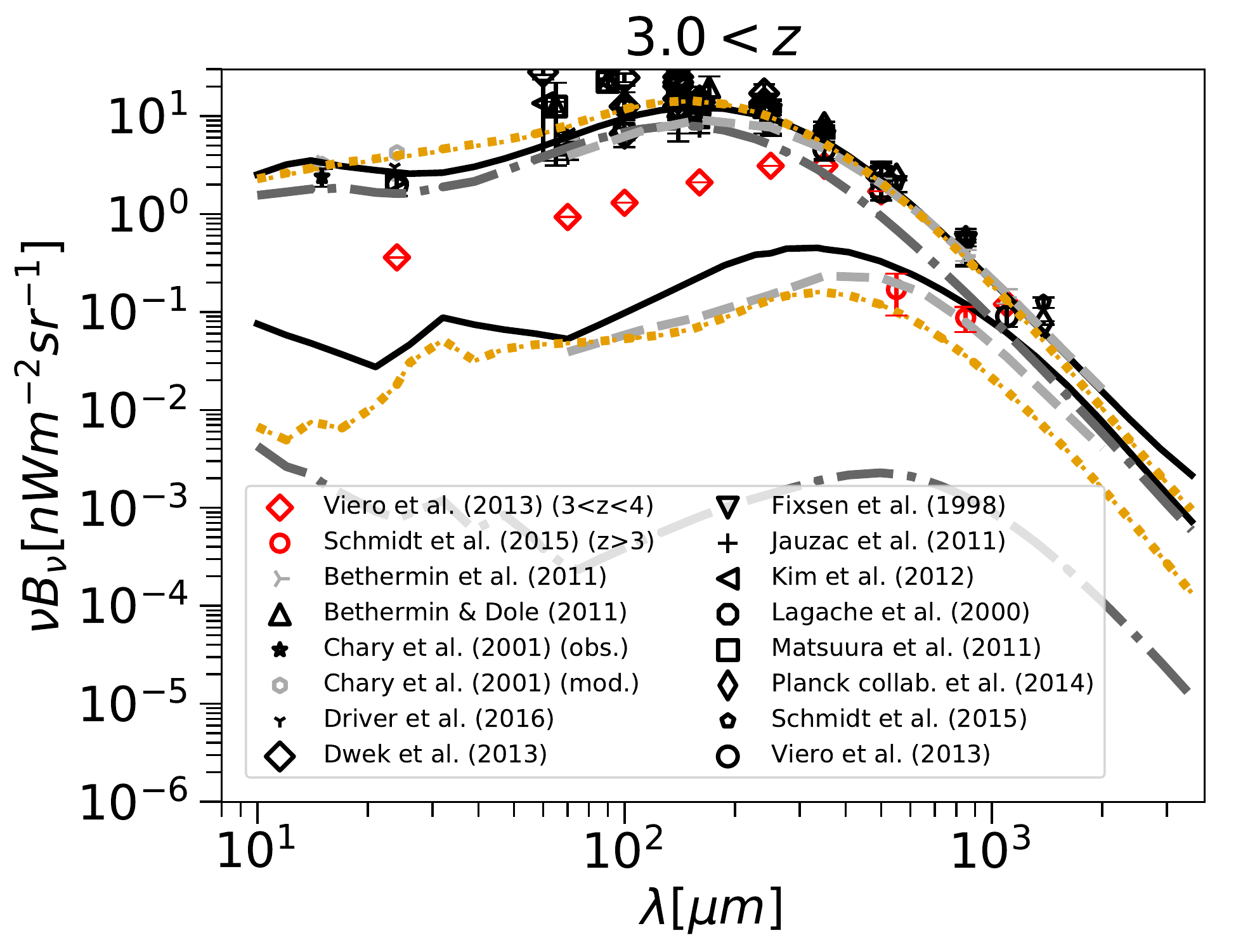}
   \caption{Extragalactic background light spectrum. Legend is the same for all four subplots.
    1) The predictions of the BM model.
    2) P2017.
    3) E2.
    4) E2 without AGN contribution.
    The four plots shown demonstrate four redshift slices.
    Upper left panel:  $0.0<z<1.0$.
    Upper right panel: $1.0<z<2.0$.
    Lower left panel:  $2.0<z<3.0$.
    Lower right panel: $3.0<z$.
    Upper group of curves are the EBL SEDs for the whole redshift range.
    The following data are shown on the plot.
    \textbf{Upper left and right panels.}     
     \cite{2012A&A...542A..58B}, 
     \cite{2013ApJ...779...32V}~-- Hershel. 
     \cite{2011A&A...525A..52J}~-- Spitzer. 
     \cite{2015MNRAS.446.2696S}~-- Planck. 
    \textbf{Lower left panel.}
     \cite{2012A&A...542A..58B}, 
     \cite{2013ApJ...779...32V}~-- Hershel. 
     \cite{2015MNRAS.446.2696S}~-- Planck. 
    \textbf{Lower right panel.}
     \cite{2013ApJ...779...32V}~-- Hershel, $3<z<4$. 
     \cite{2015MNRAS.446.2696S}~-- Planck, $z>3$. 
     \cite{2011A&A...529A...4B}~-- Model predictions. 
     \cite{2011arXiv1102.1827B}~-- Observational data from different instruments. See original paper for details. 
     \cite{2001ApJ...556..562C}~-- Observational data and model predictions. 
     \cite{2016ApJ...827..108D}~-- Spitzer and Hershel. 
     \cite{2013APh....43..112D}~-- Observational data from different instruments. See original paper for details. 
     \cite{1998ApJ...508..123F}, 
     \cite{2000A&A...355...17L}~-- COBE/FIRAS. 
     \cite{2011A&A...525A..52J}~-- Spitzer. 
     \cite{2012JPSJ...81b4101K}~-- Akari and Spitzer. 
     \cite{2011ApJ...737....2M}~-- AKARI. 
     \cite{2014A&A...571A..30P}, 
     \cite{2015MNRAS.446.2696S}~-- Planck. 
     \cite{2013ApJ...779...32V}~-- Hershel. 
    }
    \label{fig:nubnu_vs_lambda_redshifts}
  \end{figure*}

  In this section we consider the dependence of the Extragalactic Background Light on the wavelength $\lambda$ (SED).
  The EBL SEDs for four redshift intervals are shown in Fig.~\ref{fig:nubnu_vs_lambda_redshifts}.
  All the models considered show good agreement with observational data and with each other on relatively low redshift ($z<2$).
  The same is true for the total EBL SED.
  As was already mentioned above, active galactic nuclei have noticeable contribution only on relatively short wavelengths, and 
  it is most prominent at the $0<z<1$ redshift interval.
  
  On wavelengths shorter than 100$\mathrm{\mu m}$ Zodiacal light exceeds the Extragalactic Background and only indirect estimates via number counts can be made.
  See, e.g. \cite{2016ApJ...827....6S}.
      
 \subsection{The confusion noise}
 \label{subsec:confusion_noise}
 
 There is some confusion in the term ``confusion'' that different authors define in various ways.
 We used the following estimation criterions.
 
 This first criterion can be defined as the minimal completeness of detection of sources with flux greater than $S_{lim}$.
 It is defined through a fraction of sources that are lost in the detection process when a neighbor with a flux greater than $S_{lim}$ is at an angular distance
 that makes the separation impossible.
 
 Frequently the following formula is used \citep{2003ApJ...585..617D}:
 \begin{equation}
  N_{SDC}=-\dfrac{\log(1-P(<\theta_{min}))}{\pi k^2\theta_{FW}^2}
  \label{confusion:sdc}
 \end{equation}
 Here the probability $P=0.1$ and $k=0.8$, while $\theta_{FW}$ is a full width at half magnitude of the beam profile.
 
 The second criterion is called photometric and is calculated in the following way \citep{2003ApJ...585..617D}.
 First, the response amplitude $x$ from the source with flux $S$ and coordinates $\theta$, $\phi$ is defined:
 
 \begin{equation}
  x=Sf(\theta,\phi)
  \label{confusion:phot_xs}
 \end{equation}
 Where $f(\theta,\phi)$ is the two-dimensional shape of the beam profile.
 The average number of responses $R(x)$ with amplitudes from $x$ to $x+dx$ from sources in the element of the beam profile
 $d\Omega$ with coordinates $(\theta,\phi)$, where $d\Omega=2\pi\theta d\theta d\phi$ can be found in the following way:
 \begin{equation}
   R(x)dx=\int_{\Omega}\dfrac{dN}{dS}dSd\Omega
  \label{confusion:phot_rx}
 \end{equation}
 The dispersion of the measurement in the  beam due to the extragalactic sources with flux lower than $S_{lim}$ is expressed as:
 \begin{equation}
  \sigma_c^2=\int_0^{x_{lim}}x^2R(x)dx
  \label{confusion:phot_sigma_c2_1}
 \end{equation}
 where $x_{lim}=S_{lim}f(\theta,\phi)$ is the upper limit of response on large fluxes.
 \cite{2003ApJ...585..617D} rewrite this in the following way
 \begin{equation}
  \sigma_c^2=\int f^2(\theta,\phi)d\theta d\phi=\int_0^{S_{lim}}=S^2\dfrac{dN}{dS}dS
  \label{confusion:phot_sigma_c2_2}
 \end{equation}
 Here $dN/dS$ is the differential number counts expressed in $Jy^{-1}Sr^{-1}$, $\sigma_c$ is the confusion noise,
 $S_{lim}$ the confusion limit.
 As a next step the photometric criterion $q$ must be set.
 The common choice is 3 or 5. Then the following equation must be solved.
 \begin{equation}
  q=\dfrac{S_{lim}}{\sigma_{c phot}(S_{lim})}
  \label{confusion:phot_q}
 \end{equation}
  
 The third criterion is called `Probability of Deflection'~-- $P(D)$.
 It is calculated in the following way (see, e.g. ~\citep{2010MNRAS.409..109G}).
 The average density of sources per solid angle with flux in the interval from $x$ to $x+dx$.
 \begin{equation}
  R(x)dx=\int_\Omega\dfrac{dN}{dS}\left(\dfrac{x}{b}\right)b^{-1}d\Omega dx
  \label{confusion:p_d_rx}
 \end{equation}
 where $b$ is a beam function.
 The probability distribution function for a single pixel is:
 \begin{equation}
  P(D)=F_\omega^{-1}\left[\exp\left(\int_0^\infty R(x)\exp(i\omega x)dx-\int_0^\infty R(x)dx\right)\right]
  \label{confusion:p_d_fourier}
 \end{equation} 
 where $F_\omega^{-1}$ is the inverse Fourier transform.

 Different parameters can be used to quantify the confusion limit.
 The authors of \cite{2011A&A...529A...4B} have used, for example, an interquartile divided by $1.349$.
 
 Sometimes the confusion noise is defined as flux at which the number of sources reaches certain amount per beam area.
 Usually 1/20, 1/30 or 1/40 value is chosen.

\begin{figure*}[ht!]
  \center
  \includegraphics[width=\columnwidth]{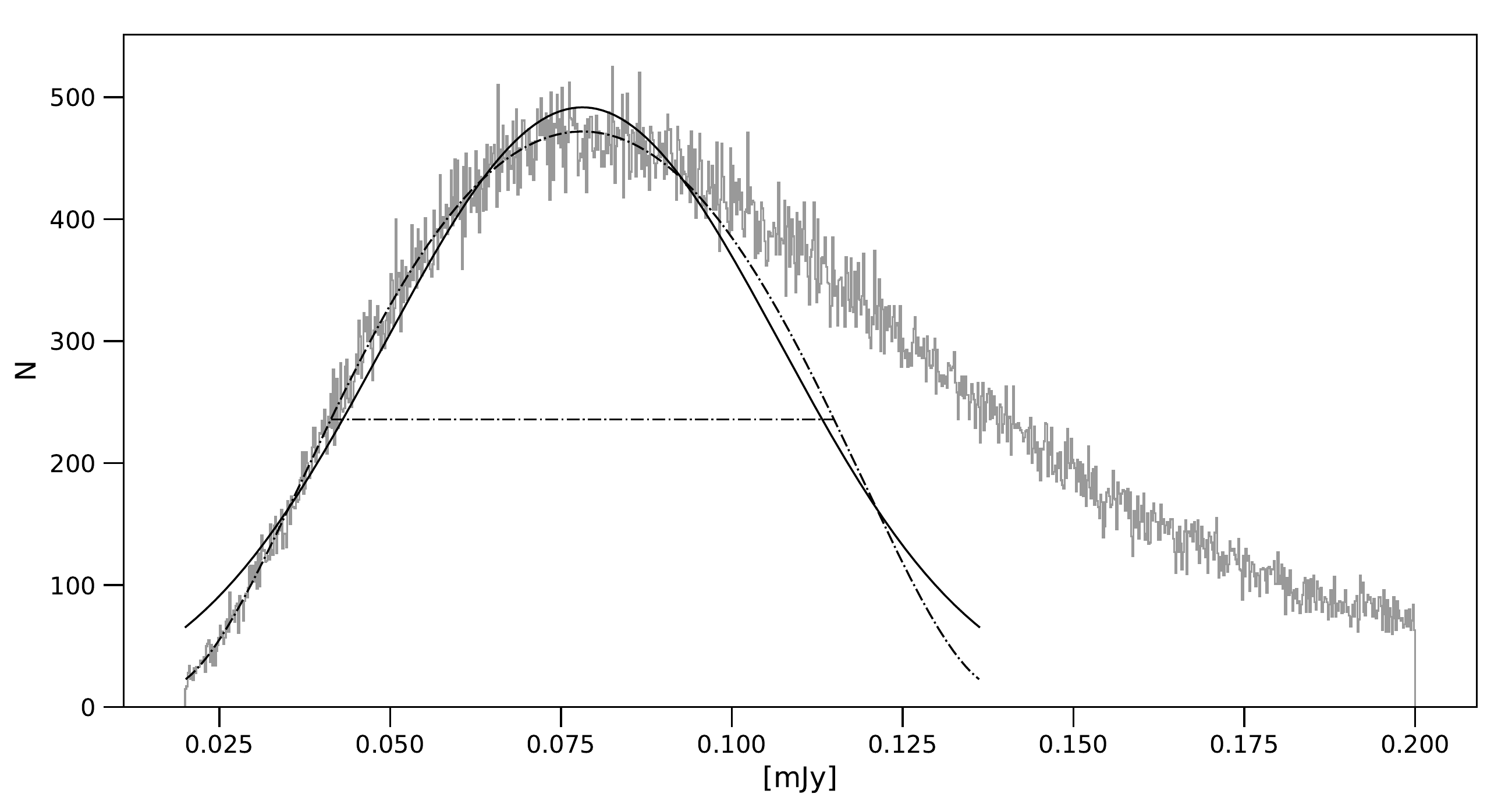}
  \caption{Histogram of the pixel flux distribution of the model map for a telescope with $10\mathrm{m}$ mirror at 850$\mathrm{\mu m}$ wavelength.
           Solid line~-- the Gaussian approximation of the negative part.
           Dot-dashed line~-- the polynomial approximation of the negative side of the histogram.
           }
  \label{fig:pixel_histogram}
 \end{figure*}

\begin{figure*}[ht!]
  \centering
  \includegraphics[width=\columnwidth]{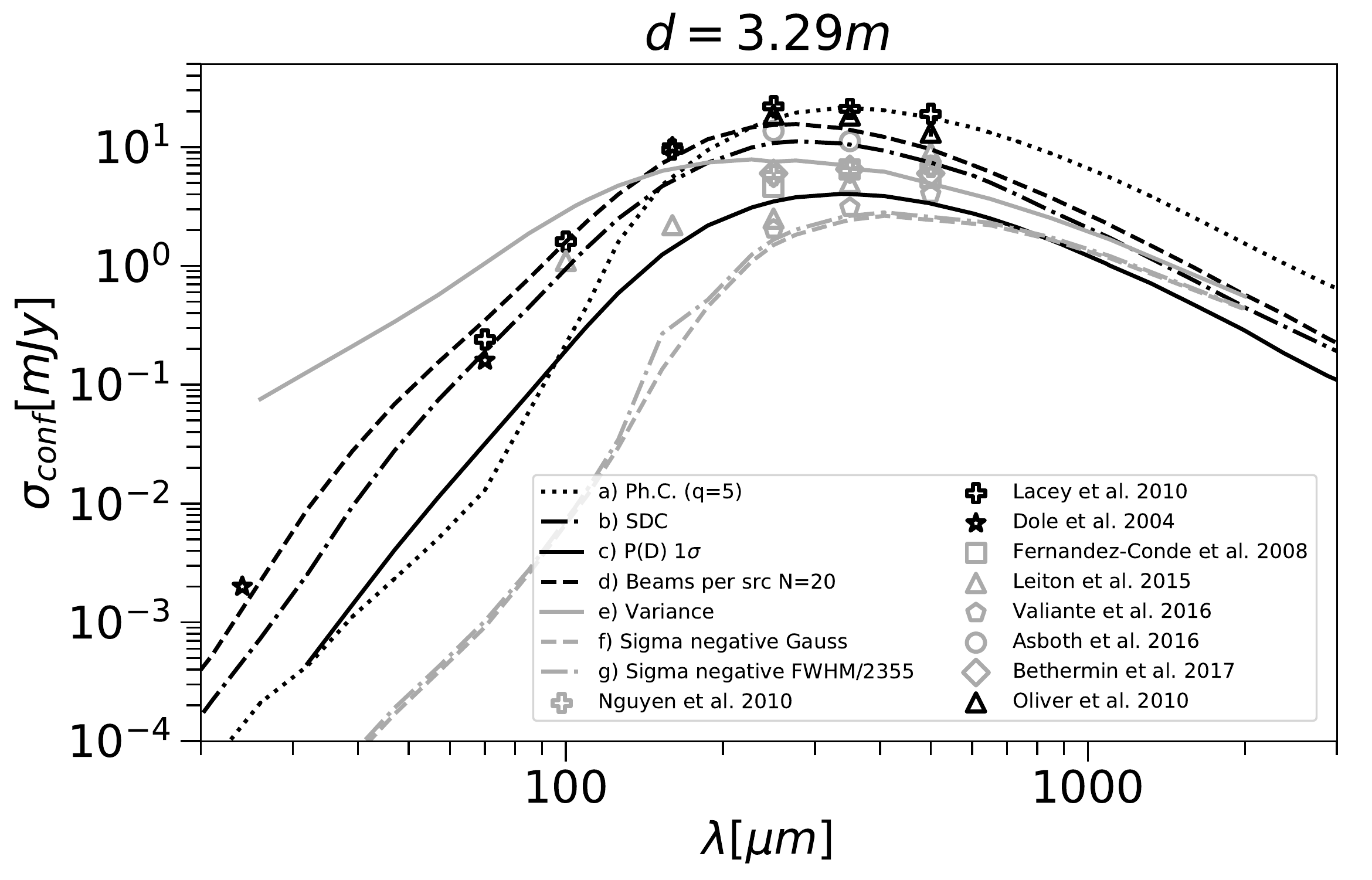}
  \caption{The dependence of the confusion noise on the wavelength for the E2 model estimated with different approaches.
           The diameter of the main mirror $d=3.29\mathrm{m}$ (Hershel).
           Black lines and markers show estimates based on source counts, gray ones show estimates based on analysis of maps.
           a) Ph.C~-- photometrical criterion, parameter $q=5$.
           b) Source Density Criterion.
           c) P(D)~-- probability of deflection criterion. Value for 1$\sigma$ is given.
           d) The number of beams per source (N=20) with flux greater than $S_\nu$.
           e) Simple pixel variance.
           f) Dispersion of the Gaussian fitted to the left part of the pixel histogram.
           g) $FWHM/2.355$ of the left side of the pixel histogram.
           The following estimations from the literature are overplotted:
           \cite{2010A&A...518L...5N}~--      
           estimation of confusion noise from observational maps.
           \cite{2010MNRAS.405....2L}~---     
           $N=20$ beams per source.
           \cite{2004ApJS..154...93D}~---     
           Source Density Criterion (SDC).
           \cite{2008A&A...481..885F}~--- 
           the flux variance in pixels of the map.
           \cite{2015A&A...579A..93L}~--- 
           the confusion noise is defined as the flux of 68\% completeness of the map.
           \cite{2016MNRAS.462.3146V}~---     
           $\sigma$ of the Gaussian function fitted to the left side of the diagram.
           \cite{2016MNRAS.462.1989A}~---     
           variance of the flux in the pixels of the map.
           \cite{2017A&A...607A..89B}~--- 
           confusion noise from analysis of model maps.
           \cite{2010A&A...518L..21O}~--- 
           $N=20$ beams per source.           
  }
  \label{fig:fig_sigma_vs_lambda_d_3_29}
 \end{figure*}

\begin{figure*}[ht!]
  \centering
  \includegraphics[width=\columnwidth]{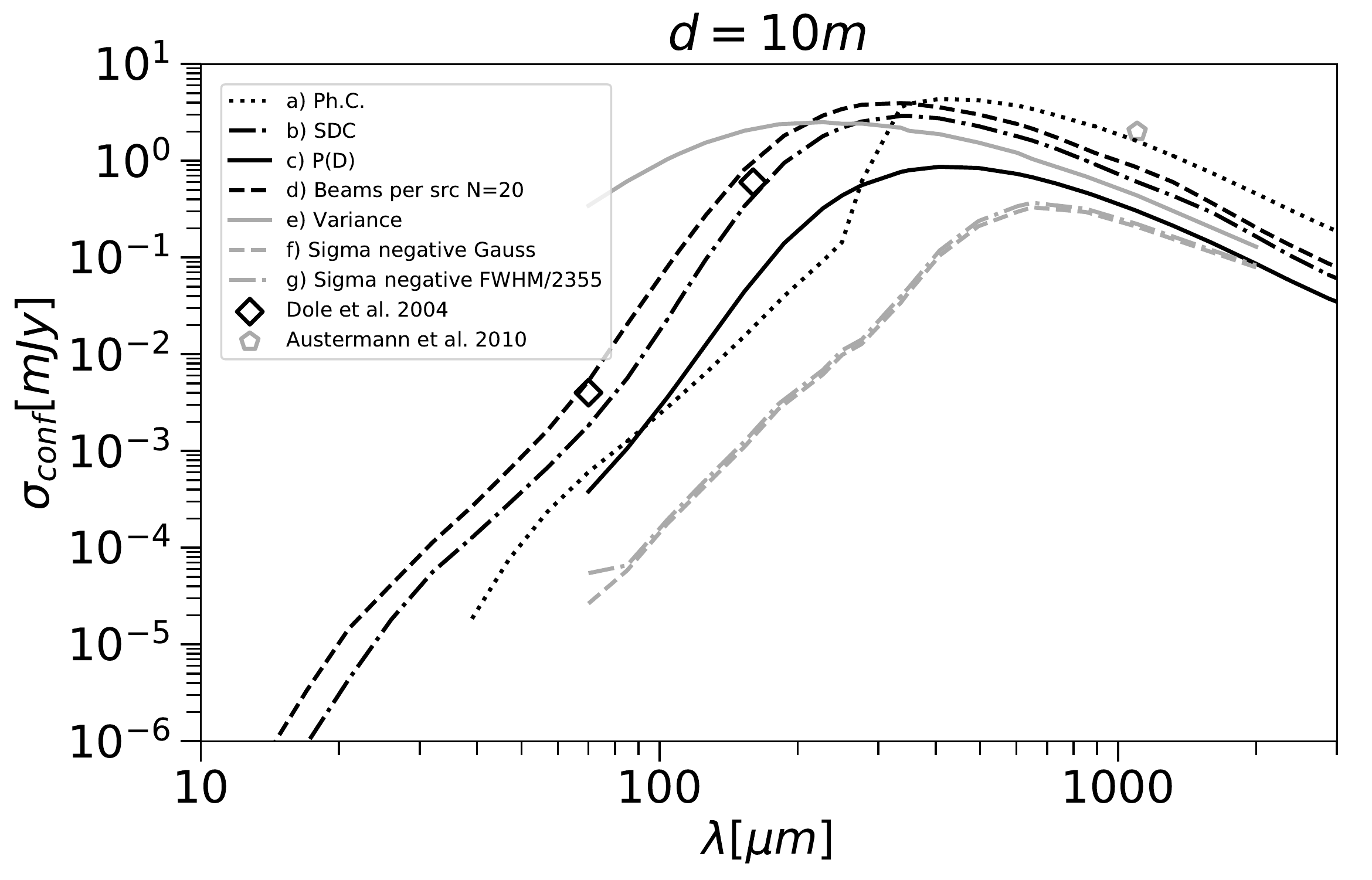}
  \caption{The dependence of the confusion noise on the wavelength for the E2 model estimated with different approaches.
           The diameter of the main mirror $d=10.0\mathrm{m}$ (Hershel).
  The legend is identical to the legend of Fig.~\ref{fig:fig_sigma_vs_lambda_d_3_29} except the observational estimations.  
   \cite{2004ApJS..154...93D}~-- 
    Source Density Criterion.
   \cite{2010MNRAS.401..160A}~-- 
    $N=30$ beams per source.
  }
  \label{fig:fig_sigma_vs_lambda_d_10_00}
 \end{figure*}

 All the methods described above are based solely on the information about the number counts.
 Other methods are based on the map analysis.
 The simplest possible approach is to measure the variance of flux in pixels of a map, see e.g., \cite{2008A&A...481..885F}.
 But the validity of such an approach is questionable because the flux distribution in pixels is non-Gaussian
 \citep{2010A&A...518L...5N,2009ApJ...707.1729M,2015A&A...579A..93L} that can be clearly seen in Fig.~\ref{fig:pixel_histogram}. 
 In some papers (see, e.g., \cite{2009ApJ...707.1729M}) the confusion noise is defined as  $\sigma$ of the Gaussian fitted to the left side of the pixel 
 histogram.
 But at shortest wavelengths if observations are performed with high angular resolution (e.g. $\lambda=250\mathrm{\mu m}$ and $d=10\mathrm{m}$) the left side of the histogram
 also shows deviation from the Gaussian form.
 In Figs.~\ref{fig:fig_sigma_vs_lambda_d_3_29} and \ref{fig:fig_sigma_vs_lambda_d_10_00} we give estimations of confusion noise as $\sigma$ of the fitted Gaussian to the left side of the histogram, and also its FWHM/3.

 Figures~\ref{fig:fig_sigma_vs_lambda_d_3_29} and \ref{fig:fig_sigma_vs_lambda_d_10_00} show the dependence of the confusion noise on the wavelength
 for two instruments with the diameter of the main mirror $d=3.29\mathrm{m}$ (Hershel) and $d=10\mathrm{m}$ (Millimetron).

 Because we consider space telescopes we supposed the image quality to be diffractional, so the confusion noise depends on the angular resolution and the 
 shape of the number counts curve.
 The angular resolution depends on the diameter of the main mirror of the telescope and on the wavelength of observations.
 So the general trend of increasing of the confusion noise with the decrease of the diameter of the main mirror and with increase of the wavelength
 is overlayed by the effect of changing of the shape of the number counts curves.
 For various diameters of the main mirror the confusion noise curves have the same shape: they peak at certain wavelength and
 decline towards shorter and longer wavelengths.
 
\section{Conclusions}
\label{sec:conclusions}

In this paper we constructed a model of extragalactic background light and compared it with three other models and with observational data.
The primary model, referred to as E2, was based on the eGALICS simulation~\citep{2015A&A...575A..32C,2015A&A...575A..33C}.
 With help of GRASIL and CHE\_EVO code \citep{1998ApJ...509..103S} we created the library of SEDs of discs and bulges for a given parameter grid.
 Each disc or bulge in the eGALICS simulation was assigned an object ID with closest parameters in the SED catalog.
 The inclination of each galaxy was set random, the AGN contribution and gravitational lensing were taken into account.

 We compared this model with a widely used in literature model of~\cite{2011A&A...529A...4B}, referred to as BM in text. We have reproduced source counts using the recipes published in~\cite{2011A&A...529A...4B}. We also compared the results with our previously published model~\citep{2017AstL...43..644P} (P2017).
  
 With these models we calculated differential number counts, extragalactic background light SEDs for various bins in redshift,
 the dependence of the confusion noise on wavelength and telescope diameter.
 Models BM, P2017, E2 demonstrate fair agreement with each other and observational data in case such data is available.
 It is argued that deviation between predicted number counts on large wavelengths seems to be caused by the resolution effects.
 
 We have analysed the number counts of sources on a 2D flux-redshift plane.
 There is a trend of increasing of contribution of distant galaxies on longer wavelengths.
 
 All models reproduce the total background SED fairly well.
 The same is true for relatively small redshifts: $0<z<1$ and $1<z<2$.

 We used different methods to estimate confusion limits for telescopes with diameters of $3.29\mathrm{m}$ and $10\mathrm{m}$ on wavelengths 70--2000$\mathrm{\mu m}$.
 It was discussed how the shape of the curve of the source number counts affects the confusion noise estimates.
 It is shown that there is a ``confusion'' in estimation of the confusion noise.
 Various criterions should be compared with caution and should be chosen according to the particular need.
 
 As regards the specific recommendations on observations, it is too early to make them at present.
 First, the parameters of the detectors still await the final approval; the same is true for the ultimate list of scientific tasks.
 Second, the approaches that allow the confusion noise to be overcome using photometric data at various wavelengths are currently being actively developed by various authors.
 Their availability will also affect the strategy of future observations.

\section{Acknowledgements}
 This research has made use of the Tool for OPerations on catalogs And Tables (TOPCAT; Taylor 2005): www.starlink.ac.uk/topcat/;  
 
 This research made use of SciPy and NumPy~\cite{scipy_reference}, and the AstroML~\cite{astroML} package.

 The work of A.A. Ermash and S.V. Pilipenko was supported by project N01-2018 of the Program ``New Scientific Groups of the Lebedev Physical Institute
 of the Russian Academy of Sciences.'' The work of V.N. Lukash was supported by the Russian Foundation for Basic Research (project no. 19-02-00199).
 This publication was supported Project KP 19-270 of the Russian Academy of Sciences.

\bibliographystyle{unsrt} 
\bibliography{ermash_pilipenko_lukash}

\end{document}